\documentclass[preview,11pt]{elsarticle}
\usepackage{amsmath,color,graphicx,amssymb}
\usepackage{natbib}
\usepackage{amsbsy}
\usepackage[]{ar}
\usepackage{latexsym}
\usepackage{subfigure}
\usepackage[english]{babel}
\usepackage{ifsym}
\usepackage{bbding}
\usepackage{wasysym}
\usepackage{epsfig}
\usepackage{epsf,psfig}
\usepackage{epstopdf}
\usepackage{color}
\usepackage{multirow}
\usepackage[margin=1in]{geometry}
\usepackage{threeparttable}
\usepackage{booktabs,array}
\usepackage{array}
\usepackage{float}

\definecolor{gray}{rgb}{0.5,0.5,0.5}
\definecolor{purple}{rgb}{0.57,0,0.86}

\newcommand\mb{\mathbf}
\newcommand\bsy{\boldsymbol}
\newcommand\bnabla{\boldsymbol{\nabla}}
\newcommand\mc{\mathcal}

\def\beq{ \begin{equation}}
\def\eeq{\end{equation}}
\def\beqar{ \begin{eqnarray} }
\def\eeqar{ \end{eqnarray} }

\def\vK{von K\'{a}rm\'{a}n }

\journal{Computers and Fluids}

\begin{document}

\begin{frontmatter}



\title{Unsteady Three-Dimensional Boundary Element Method for Self-Propelled Bio-Inspired Locomotion}


\author[label1]{K.~W. Moored}
\address[label1]{Department of Mechanical Engineering and Mechanics, Lehigh University, Bethlehem, PA 18015, USA}



\begin{abstract}
An unsteady three-dimensional boundary element method is developed to provide fast calculations of biological and bio-inspired self-propelled locomotion.  The approach uniquely combines an unsteady three-dimensional boundary element method, a boundary layer solver and self-propelled equations of motion.  This novel implementation allows for the self-propelled speed, power, efficiency and economy to be accurately calculated.  A Dirichlet formulation is used with a combination of constant strength source and doublet elements to represent a deforming body with a nonlinearly deforming wake.  The wake elements are desingularized to numerically stabilize the evolution of the wake vorticity.  Weak coupling is used in solving the equations of motion and in the boundary layer solution.  The boundary layer solver models both laminar and turbulent behavior along the deforming body to estimate the total skin friction drag acting on the body.  The results from the method are validated with analytical solutions, computations and experiments.  Finally, a bio-inspired self-propelled undulatory fin is modeled.  The computed self-propelled speeds and wake structures agree well with previous experiments.  The computations go beyond the experiments to gain further insight into the propulsive efficiency for self-propelled undulating fins.  It is found that the undulating fin produces a time-averaged momentum jet at 76\% of the span that accelerates fluid in the streamwise direction and in turn generates thrust.  Additionally, it is discovered that high amplitude motions suppress the formation of a bifurcating momentum jet and instead form a single core jet.  Consequently, this maximizes the amount of streamwise momentum compared to the amount of wasted lateral momentum and leads to a propulsive efficiency of 78\% during self-propelled locomotion. 
\end{abstract}

\begin{keyword}
Bio-inspired propulsion \sep boundary element method \sep panel method \sep self-propelled swimming \sep unsteady flows \sep vortex dynamics 


\end{keyword}

\end{frontmatter}


\section{Introduction}

Boundary element methods (BEMs) are a class of numerical methods used to solve boundary value problems throughout physics from electromagnetics \cite[]{Kagami1984} and fracture mechanics \cite[]{Portela1992} to fluid flows at both low \cite[]{Pozrikidis2001} and high Reynolds numbers \cite[]{Basu1978}.  In high Reynolds number flows they are classically described as panel methods and have been well established in the study of aerodynamics over several decades \cite{Hess1972,Maskew1987,Katz2001}.  High Reynolds number BEMs assume that a fluid flow is incompressible, irrotational (except at singular elements) and inviscid, that is, a potential flow.  This leads to simplified forms of the continuity and momentum equations that govern the fluid flow.  Yet, unsteady BEM solutions are still rich with flow physics \cite[]{Quinn2014} and give accurate solutions at computational times that are several orders of magnitude faster than Navier-Stokes solvers \cite[]{Pan2012, Willis2006}.  

Unsteady three-dimensional BEM computations have been used by many researchers to explore both biological and bio-inspired propulsion.  The flight performance of birds \cite{Smith1996} and the swimming performance of fin whales \cite{Liu1997} and fish \cite{Cheng2001} have been examined to reveal features of high efficiency locomotion.  For example, Zhu et al. \cite{Zhu2002} found that constructive or destructive interactions can occur between the shed vorticity from finlet structures and the caudal fin of tuna and giant danio.  This can lead to enhanced thrust production or efficiency, respectively, with maximum efficiencies of 75\% being calculated.  More recently, Zhu \cite{Zhu2007} showed that spanwise and chordwise flexibility can enhance both thrust production and efficiency of a flapping wing.  The benefit of flexibility was also found to be highly dependent upon the mass ratio between the wing and the surrounding fluid environment.  Additionally, Zhu \cite{Zhu2008}, Shoele \& Zhu \cite{Shoele2009} and Shoele \& Zhu \cite{Shoele2010} determined that the flexibility of ray-finned fish caudal and pectoral fins also improved their efficiency performance and reduced the time-varying lateral forces acting on the fish.  Importantly, none of these previous studies have examined the locomotion of \textit{self-propelled} swimmers nor the \textit{free-flight} of flyers, yet these conditions are a critical feature of bio-inspired locomotion.

One complicating factor is that an inviscid BEM does not inherently calculate viscous drag.  This gives no opposing force to balance the thrust production, which leaves out a necessary ingredient for calculating a steady-state self-propelled speed.  However, viscous drag has been estimated in several other BEM studies by using a boundary layer momentum-integral approach on streamwise strips \cite{Robinson1988, Liu1999,Voutsinas2006}.  Even with a viscous drag estimate included these studies focused on fixed freestream velocity conditions.

This work describes a novel implementation for computing the self-propelled performance of biological and bio-inspired propulsors within a BEM framework.  There are three main components that must be combined to model self-propelled swimming: (1) a three-dimensional BEM fluid solver,  (2) a boundary layer solver, and (3) an equations of motion solver.  These components to the method are described in section \ref{sec:methods}.  Validation with several analytical, numerical and experimental solutions are presented in section \ref{sec:validation}.  Finally, comparison of the BEM solution with a three-dimensional self-propelled undulating fin experiment is presented in section \ref{sec:results}.  The free-swimming performance and wake structures are shown to agree well with the experiments.  Additionally, the self-propelled performance of cases that extend beyond the previous experiments are examined to provide novel physical insight into the self-propulsion of three-dimensional ray-inspired fins.

\section{Computational Methods} \label{sec:methods}

\subsection{Governing Equations and Boundary Conditions}
\begin{figure}[h!]
	\centering
		\includegraphics[width=0.85\textwidth]{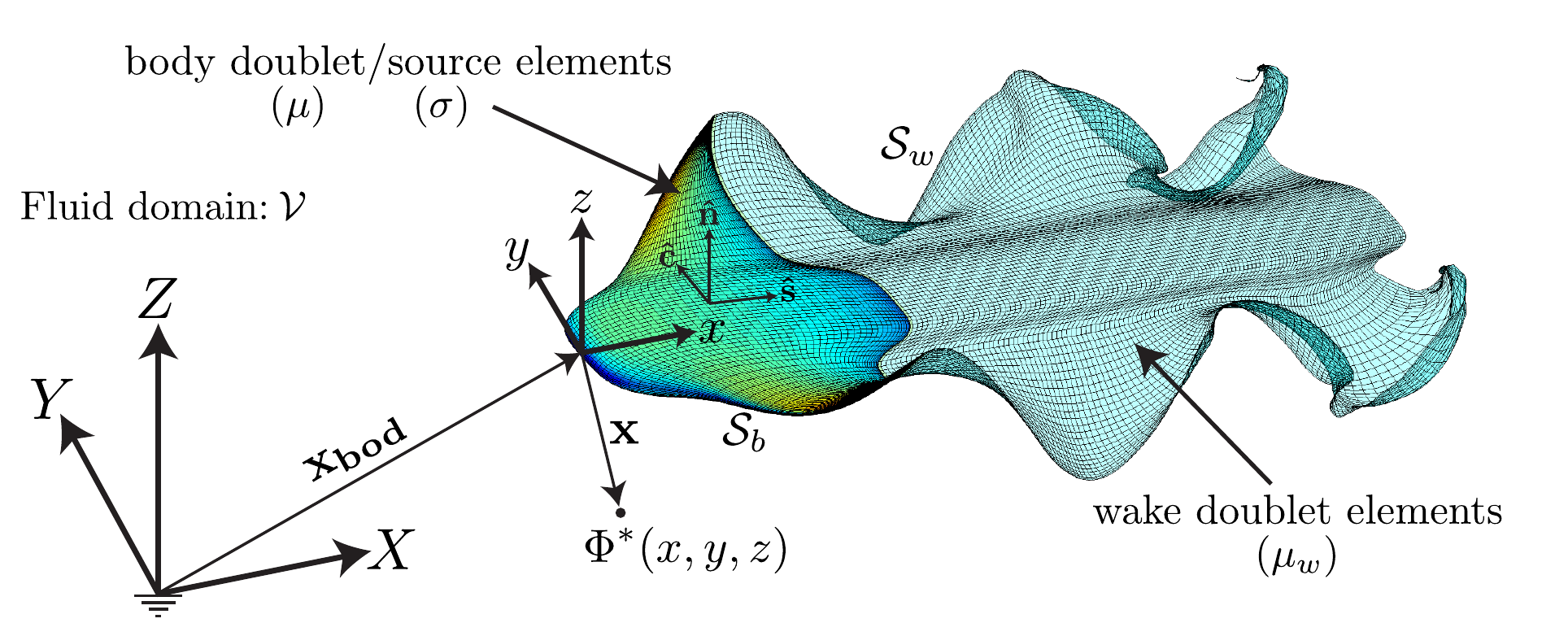}
	\caption{The inertial reference frame fixed to the undisturbed fluid is denoted by $(X,Y,Z)$ while the body-fixed reference frame is denoted by $(x,y,z)$.  The local normal, streamwise and cross-stream unit vectors are denoted by $\mathbf{\hat{n}}$, $\mathbf{\hat{s}}$, and $\mathbf{\hat{c}}$, respectively.  The body surface, $\mc{S}_b $, is layered with distributions of doublet elements of strength $\mu$ and source elements of strength $\sigma$.  The wake surface, $\mc{S}_w$, is layered with distributions of doublet elements of strength $\mu_w$.}
	\label{BEM}
\end{figure}

To model a high Reynolds number fluid flow around an self-propelled bio-inspired device or animal an unsteady three-dimensional boundary element method is employed.  The flow field is modeled as an incompressible, irrotational and inviscid flow, that is, a potential flow.  For the self-propelled problem we define the problem in an inertial frame of reference that is attached to the undisturbed fluid (denoted by $(X,Y,Z)$ is Figure \ref{BEM}).  As such the velocity field, $\mb{u}$, may be defined everywhere as the gradient of a scalar velocity potential, 
\begin{equation} \label{VelPot}
\mb{u} = \bsy{\nabla} \Phi^*,
\end{equation}

\noindent where $\Phi^*$ is defined in the inertial frame of reference and it is known as the perturbation potential.  The pressure field within this fluid can be calculated from the unsteady Bernoulli equation,
\begin{equation} \label{BernI}
P(X,Y,Z,t) = -\rho \frac{\partial \Phi^*}{\partial t}\bigg|_{inertial} - \rho \frac{\left(\bnabla \Phi^* \right)^2}{2},
\end{equation}

\noindent which is formulated here in the inertial frame where the reference pressure $P_\infty = 0$ and the perturbation potential at infinity is zero.  The time derivative of the perturbation potential for a point on the surface of the body is then calculated by using a body-fixed Lagrangian frame (denoted by $(x,\, y,\, z)$ in Figure \ref{BEM}) \cite{Cheng2001,Willis2007,Pan2012}, that is,
\begin{align} \label{BernBody}
 P(x,y,z,t) = -\rho \frac{\partial \Phi^*}{\partial t}\bigg|_{body} + \rho \left(\mb{u_{rel} + \mb{U_0}} \right) \cdot \bnabla \Phi^*- \rho \frac{\left(\bnabla \Phi^* \right)^2}{2}.
\end{align}

\noindent The translational velocity of a body-fixed frame of reference is $\mb{U_0}$ while the relative velocity of a point on the surface of the body to the body-fixed reference frame is $\mb{u_{rel}}$.  Once the perturbation potential is known, then the pressure on the body surface may be found and the forces can be calculated by integrating the pressure and shear stress acting on the body.
\begin{equation} \label{Forces}
\mb{F}\left(x,y,z,t \right) =  \int_{\mc{S}_b } \left(-P\; \mb{\hat{n}} + \tau \; \mb{\hat{s}} \right)\; d\mc{S} 
\end{equation}

\noindent The body surface is denoted as $\mc{S}_b $, the outward normal vector from the body surface is $\mb{\hat{n}}$ and the tangential vector along the body surface in the streamwise direction is $\mb{\hat{s}}$.  This inviscid formulation is coupled to a viscous boundary layer solver described in section \ref{BLSolve}, which estimates the shear stress, $\tau$, acting on the body in the streamwise direction produced by the outer potential flow.  Note that the shear stress acting in the cross-stream direction is not accounted for in the viscous boundary layer solver and is therefore not present in equation (\ref{Forces}). The problem is then reduced to solving for the perturbation potential throughout the fluid, which is governed by Laplace's equation, \begin{equation} \label{LaplaceI}
\nabla^2 \Phi^* = 0.
\end{equation}

\noindent The boundary conditions that must be satisfied for an inviscid fluid are that there is no fluid flux through the body surface and that the flow disturbances caused by the body must decay far away,
\begin{align}  \label{NoFlux}
&\mb{n} \cdot \bsy{\nabla} \Phi^* = \mb{n} \cdot \left( \mb{u_{rel} + \mb{U_0} } \right) \quad \quad \mbox{on } \mc{S}_b \\
&\bsy{\nabla} \Phi^* \bigg|_{|\mb{x}| \rightarrow \infty} = 0 \quad \quad \mbox{on } \mc{S}_\infty
\end{align}

\noindent where $\mc{S}_\infty$ is the surface at infinity bounding the fluid and $\mb{x} = \left[ x, \, y, \,  z\right]^T$ is measured from the body-fixed frame of reference.

\subsection{Boundary Integral Equation}
A general solution to Laplace's equation for the potential anywhere within the fluid domain, $\mathcal{V}$, can be determined.  This is done by considering the potential response at a point $\mb{x} = [x,\, y, \, z ]^T$ due to a source located at $\mb{x_0} = [x_0,\, y_0, \, z_0 ]^T$, that is, the infinite space Green's function which satisfies the Poisson equation with homogeneous far-field boundary conditions \cite{Haberman2004}.  The infinite space Green's function in three dimensions is then,
\begin{align}
G(\mb{x};\mb{x_0}) = - \frac{1}{4 \pi r}, \quad \quad \mbox{where  } r = |\mb{x} - \mb{x_0} |.
\end{align}

By invoking Green's formula twice with respect to the perturbation potential within the fluid volume, the internal perturbation potential (the volume enclosed by $\mc{S}_b$) and the Green's function and then adding the results, the following boundary integral equations (BIEs) for the internal or external perturbation potential are derived,
\begin{align} \label{IntPertPot}
\Phi_i^*(\mb{x}) & = \oiint_{\mc{S}_b} \left[ \sigma(\mb{x_0}) \; G(\mb{x};\mb{x_0}) - \mu(\mb{x_0}) \; \mb{\hat{n}} \cdot \bsy{\nabla} G(\mb{x};\mb{x_0})\right] \; d\mc{S}_0 -\oiint_{\mc{S}_w} \mu_w(\mb{x_0}) \; \mb{\hat{n}} \cdot \bsy{\nabla} G(\mb{x};\mb{x_0}) \; d\mc{S}_0  \\ 
\Phi^*(\mb{x}) & = \oiint_{\mc{S}_b} \left[ \sigma(\mb{x_0}) \; G(\mb{x};\mb{x_0}) - \mu(\mb{x_0}) \; \mb{\hat{n}} \cdot \bsy{\nabla} G(\mb{x};\mb{x_0})\right] \; d\mc{S}_0 -\oiint_{\mc{S}_w} \mu_w(\mb{x_0}) \; \mb{\hat{n}} \cdot \bsy{\nabla} G(\mb{x};\mb{x_0}) \; d\mc{S}_0  \\
& \mbox{where,} \nonumber \\
\sigma(\mb{x_0})  & = \mb{\hat{n}} \cdot \bsy{\nabla} \left(\Phi^* -  \Phi_i^*\right) \label{Source} \\
-\mu(\mb{x_0}) & = \Phi^* - \Phi_i^* \\
-\mu_w(\mb{x_0}) & = \Phi_+^* - \Phi_-^*
\end{align}

\noindent During the derivation of the boundary integral equation the source and observation locations switch roles.  Note that when this happens $G(\mb{x_0};\mb{x}) = G(\mb{x};\mb{x_0})$ due to the reciprocity of the Green's function, however, $\bsy{\nabla} G(\mb{x_0};\mb{x}) = -\bsy{\nabla} G(\mb{x};\mb{x_0})$.   The surface boundary integral is broken up into an integral over the body boundary, the wake boundary and the far-field boundary.  To formulate the problem in an inertial frame of reference attached to the undisturbed fluid the contribution to the potential from the farfield boundary is set to zero, that is $\Phi_\infty = 0$.  The potential jump $\mu(\mb{x_0})$ represents the strength of a dipole or doublet while the jump in the normal derivative of the potential $\sigma(\mb{x_0})$ represents the strength of a source.  The wake boundary in the limit as it becomes infinitesimally thin has a continuous normal derivative of the potential so it does not support a source distribution.  The local potential jump between the top and bottom surface of the wake is represented by $\mu_w(\mb{x_0})$, where $\Phi^*_+$ is the potential above the wake and $\Phi^*_-$ is the potential below the wake. 

The general solution to the potential flow problem in a fluid domain $\mc{V}$ is now reduced to finding a distribution of sources and doublets on the boundaries $\mc{S}_b$ and $\mc{S}_w$ that satisfy the boundary conditions.  Since the continuity and thus Laplace's equation is time-independent, all of the time-dependency comes from the unsteady Bernoulli equation and the time-dependent boundary conditions.  

\subsection{Enforcement of the Boundary Conditions}
The elementary solutions of the doublet and source both implicitly satisfy the far-field boundary condition. In this paper, the no-flux condition on the body is satisfied with an indirect Dirichlet formulation since it saves computational effort as compared to the Neumann formulation \cite[]{Katz2001}.  For the Dirichlet problem, we set the internal potential to a constant, which in our case is selected to be zero
\begin{align} \label{Dirichlet}
\Phi_i^* = 0.
\end{align}

To enforce this condition the BIE must be satisfied with the left-hand side equal to zero.  This condition also leads to the simplification of the source and doublet strength equations, which become 
\begin{align}
-\mu & = \Phi^* \\
\sigma & = \bsy{\nabla} \Phi^* \cdot \mb{n} = \left(  \mb{u_{rel} + \mb{U_0} } \right) \cdot \mb{n}.
\end{align}

The perturbation velocity on the surface of the body is simplified and can be found by a local differentiation of the perturbation velocity potential,
\begin{align}
\mb{u_b} = \bnabla \Phi_b^* = \frac{\partial \Phi*}{\partial s} \; \mb{\hat{s}} + \frac{\partial \Phi*}{\partial c} \; \mb{\hat{c}} + \frac{\partial \Phi*}{\partial n} \; \mb{\hat{n}} = - \frac{\partial \mu}{\partial s} \; \mb{\hat{s}}  - \frac{\partial \mu}{\partial c} \; \mb{\hat{c}} +  \sigma \; \mb{\hat{n}},
\end{align}

\noindent where $\mb{\hat{c}}$ is the tangential vector along the surface in the cross-stream direction.  The pressure over the body can also be found in terms of the boundary element strengths, 
\begin{align} \label{BernBody}
 P_b(x,y,z,t) = \rho \frac{\partial \mu}{\partial t}\bigg|_{body} + \rho \left(\mb{u_{rel} + \mb{U_0}} \right) \cdot \mb{u_b} - \rho \frac{\mb{u_b} ^2}{2}.
\end{align}

Now, the problem is reduced to finding a distribution of sources and doublets that solve the BIE when the Dirichlet condition is applied.  

\subsection{Numerical Solution}
To numerically solve this problem, the boundaries are discretized into constant strength quadrilateral boundary elements over the body and wake.  Then the boundary integral equation (eq. (\ref{IntPertPot})) with the Dirichlet condition substituted (eq. (\ref{Dirichlet})) can be discretized into the summation over a series of boundary elements,  
\begin{align} \label{Discretized}
 \sum^{N_b}_{j = 1}  B_{ij} \sigma_j & + 
 \sum^{N_b}_{j = 1}  C_{ij} \mu_j 
 +  \sum^{N_w}_{k = 1}  C_{w,ik}  \mu_{w,k}   = 0 \\
& \mbox{with,} \nonumber \\
B_{ij}  & = -\frac{ 1}{4 \pi}  \int_{ele}  \frac{1} {| \mb{r_{ij}} |} \; \mbox{d}\mc{S}_0 \\
C_{ij} & =  -\frac{ 1}{4 \pi}  \int_{ele}\frac{\mb{\hat{n}}  \cdot  \mb{r_{ij}} }{|\mb{r_{ij}}|^3} \; \mbox{d}\mc{S}_0  \\
C_{w,ik} & =  -\frac{ 1}{4 \pi}  \int_{ele}\frac{\mb{\hat{n}}  \cdot  \mb{r_{ik}} }{|\mb{r_{ik}}|^3} \; \mbox{d}\mc{S}_0 \\
\mbox{and}  \quad \mb{r_{ij}} =& \mb{x_i} - \mb{x_{0,j}},  \quad \mb{r_{ik}} = \mb{x_i} - \mb{x_{0,k}}  
\end{align} 

\noindent where $N_b$ is the number of body elements,  $N_w$ is the number of wake elements, $d\mathcal{S}_0$ is the differential area of a boundary element, $\mb{x_i}$ is the vector denoting the position of the $i^{th}$ collocation point, $\mb{x_{0,j}}$ is the vector denoting the position of a differential area of the $j^{th}$ element and $\mb{x_{0,k}}$ is the vector denoting the position of a differential area of the $k^{th}$ element.  Explicit solutions for the influence integrals over the elements can be found in Katz \cite{Katz2001}.  Equations (\ref{Discretized}) must be satisfied at every point within the boundary $\mc{S}_b$, which in the discretized form are satisfied at $N_b$ collocation points.  In the current method, the collocation points were located at the center of the elements but moved into the body by 15\% of the distance of the half-thickness of the body at that point, along the element normal vector.  The system of equations (\ref{Discretized}), however, needs to be modified by an explicit or implicit Kutta condition to allow the support of bound circulation.  Also, for time-stepping, a wake shedding procedure must be used to satisfy Kelvin's condition.  

\subsection{Wake Model}
In this work an explicit Kutta condition is chosen for its simplicity.  A trailing-edge doublet element is the first element in the set of wake elements and it is used to enforce the Kutta condition of finite velocity at the trailing edge.  This occurs by the trailing-edge element having a strength that cancels the vorticity at the trailing-edge.  The strength of the trailing-edge element is found at each time step from the difference in strengths between the top and bottom trailing-edge doublet elements, i.e.,
\begin{align} \label{Kutta}
\mu_{w,TE} = \mu_{t,TE} - \mu_{b,TE}.
\end{align}

\noindent The orientation of the element is set to be along a line that bisects the angle of the trailing-edge, which is typical for steady flow conditions \cite[]{Katz2001}.  Traditionally, the trailing-edge element length is set to $0.3$ -- $0.5 \, U_0 \Delta t$ \cite[]{Willis2007}, where $\Delta t = 1/ (f N_{step})$ and $N_{step}$ is the number of time steps per oscillation cycle.  Here a length of $0.4 \,U_0 \Delta t$ was used since it gave good solution convergence while maintaining solution accuracy with the validation cases.  

During each time step the trailing-edge element from the previous time step is `shed' a distance $U_0 \Delta t$ downstream and it becomes the second wake element.  The strength of that wake element is the same as the strength of the previous trailing-edge element and it remains constant for all subsequent time steps.  The trailing-edge element's strength can be re-written in terms of the unknown body element strengths using equation (\ref{Kutta}) and the discretized BIE is then modified, that is, 
\begin{align} \label{BIE_Kutta}
 \sum^{N_b}_{j = 1}  A_{ij} \mu_j & = -  \sum^{N_b}_{j = 1}  B_{ij} \sigma_j  -  \sum^{N_w}_{k = 2}  C_{w,ik}  \mu_{w,k}\\
& \mbox{with,} \nonumber \\
A_{ij} & =
\begin{cases}
C_{ij} - C_{w,i1}, & \quad \mbox{j = bottom element}  \\
C_{ij} + C_{w,i1}, & \quad \mbox{j = top element} \\
C_{ij}, & \quad \mbox{otherwise} 
\end{cases} \nonumber
\end{align}

Now, the body source element strengths and wake element strengths are known.  The linear set of equations may be solved at each time step for the body doublet strengths, $\mu_j$, by an inversion of matrix $A_{ij}$.  The trailing-edge element strengths are then directly calculated from the body doublet element strengths.  In the current method, the number of unknown body doublet element strengths is reduced in half by assuming left-right symmetry for the problem and using mirror image elements to represent the left-half of the body and the wake.  

\subsection{Nonlinear Wake Deformation}\label{Wake}
The wake elements that are shed at each time step model the shedding of vorticity from the body into its wake.  These elements cannot support loads so they must be free to advect with the local velocity field.  At each time step the induced velocity at the corner points of each wake element, $\mb{u}_{w}$, is determined.  The wake element corner points are then displaced by $\Delta \mb{d} = \mb{u}_{w} \Delta t$.  Calculating the induced velocity at the wake element corner points will lead to a numerically unstable solution if the doublet elements are not desingularized.  Here we take advantage of the equivalence of a constant strength doublet element and a vortex ring around the edge of that element by using the desingularized Biot-Savart law,
\begin{align}
\mb{u(x)} = \frac{\Gamma}{4 \pi} \oint \frac{\mb{s} \times \mb{r}}{r^3 + \delta^3} \; ds,
\end{align}

\noindent to calculate the induced velocity field  \cite[]{Krasny1986}.  Here, the circulation of an element is $\Gamma = -\mu$ and the desingularization parameter, $\delta$, is a constant and a free-parameter for the method.  Provided that $\delta$ is large enough, the transfer of energy to high wavenumbers is minimized thereby preventing solution breakdown \cite[]{Zhu2002}.  The desingularization parameter mimics the effect of viscosity in a real fluid by giving each vortex ring element a core radius directly related to $\delta$.

\subsection{Lumped Wake Elements} \label{Lump}
A lumped wake element model is used to restrict the growth of the problem size as the number of wake elements increases with every time step.  The lumped elements conserve the net circulation in the wake such that Kelvin's condition still holds.  There is one lumped wake element for every trailing-edge element, with the trailing-edge elements acting as the wake element generators and the lumped elements acting as wake element absorbers in the far-field.  The strength of the lumped elements at the $n^{th}$ time step is the summation of the circulation of the lump elements at the previous time step and the absorbed elements at the $n^{th}$ time step.  
\begin{align} \label{LumpedStrength}
\Gamma_{lump}^{n} &= \Gamma_{lump}^{n-1}+ \Gamma_{w,absorbed}^{n} 
\end{align}

\noindent The lumped element corner point locations, $\mb{p}_{lump}$, are at the weighted-average locations of the absorbed elements where the weights are based on the magnitude of the increment of circulation added to the lumped element compared to the total absorbed magnitude of circulation,
\begin{align}
\mb{p}_{lump}^{n}&= \left[\frac{\Gamma_{mag}^{n-1}}{\Gamma_{mag}^{n}}\right] \mb{p}_{lump}^{n-1}+  \left[\frac{|\Gamma_{w,absorbed}^{n}|}{\Gamma_{mag}^{n}}\right] \mb{p}_{absorbed}^{n} \\
& \mbox{with,} \nonumber \\
\Gamma_{mag}^{n} &= \Gamma_{mag}^{n-1}+ |\Gamma_{w,absorbed}^{n}| \nonumber
\end{align} 

The lumped elements absorb wake elements such that the most recent $N_l$ oscillation cycles of wake elements remain to fully-resolve the near wake.  It is found that by measuring the time-averaged forces on the body, the lumped wake solution is within 1\% of the fully-resolved solution if $N_l \geq 4$ for two-dimensional flows and $N_l \geq 2$ for three-dimensional flows.  The lumped wake elements are included in the third term of equation (\ref{BIE_Kutta}).  

\subsection{Equations of Motion} \label{EoM}
To be able to calculate the self-propelled body dynamics and performance of a bio-inspired device or an animal, the equations of motion for the body must be solved.  For the results in this work, we only allow streamwise translation to be unconstrained while the other degrees of freedom undergo fully prescribed motions.  To further simplify the implementation of the unconstrained body dynamics, a loose or one-way coupling with the fluid solution is used.  That is, the forces from the solution of the BEM fluid problem are used as the driving forces in the equations of motion. The body velocity and position is then explicitly determined without sub-iterations between the fluid solver and the body dynamics solver.  Following \cite{Borazjani2008}, the loose-coupling scheme uses the body position and velocity at the current time step to explicitly solve for the position and velocity at the subsequent time step,
\begin{align} \label{EoM}
x_b^{n+1} &= x_b^{n} + \frac{1}{2}\left( U_0^{n+1} + U_0^{n} \right) \Delta t \\
U_0^{n+1} &= U_0^{n} + \frac{F_x^n}{M} \Delta t
\end{align}

\subsection{Viscous Boundary Layer Solver} \label{BLSolve}
An estimate of skin friction drag is critical to computing self-propelled body dynamics and performance as this is the main source of drag for streamlined swimmers.  Using the potential outer flow solution, the boundary layer properties and the skin friction can be estimated without an assumption on the form of a drag law acting on a body. To calculate the boundary layer properties a two-dimensional \vK momentum integral analysis is performed over streamwise strips along the body.  Some three-dimensionality is embedded into the analysis by using an effective outer flow velocity at each chordwise station, which is the magnitude of the streamwise and cross-stream tangential velocities at those locations, 
\begin{align}
|q_{outer}| = \sqrt{q_{s}^2 + q_{c}^2}
\end{align}

\noindent This analysis can be used to approximate the boundary layer displacement thickness and the skin friction drag coefficient over each boundary element based on the outer flow alone.   The outer flow is coupled to the boundary layer solution in a loose coupling, that is, there are no sub-iterations between the outer flow solution and the shape of the body based on the displacement thickness of the boundary layer.  From the outer flow solution the stagnation point on the body is determined for each strip.  Then based on the stagnation point the surface is divided into an upper and lower part where each boundary layer begins growing.  First, the method of Thwaites is used to solve for the laminar boundary layer properties, next Michel's criterion is used to determine the transition to turbulence location and finally a one-seventh-power law is used to calculate the turbulent boundary layer properties \cite[]{Robinson1988,White2006}.  

\section{Numerical Method Validation} \label{sec:validation}
To validate the current boundary element method implementation a series of analytical, numerical and experimental results are used to analyze the accuracy of the current numerical method.  The validation cases include a two-dimensional steady and unsteady flow case, a three-dimensional steady and unsteady flow case, a skin friction drag case, a self-propelled biological propulsion case and a self-propelled bio-inspired undulating fin case.

\subsection{Two-Dimensional Steady Flow}
The steady pressure distribution over a two dimensional van de Vooren airfoil is first examined.  The van de Vooren airfoil is chosen since there is an exact analytical solution for its pressure distribution  \cite[]{VandeVooren1969} and since it has a finite-angle trailing edge as opposed to the cusped trailing edge of a Joukowski airfoil.  Cusped trailing edges are numerically difficult to obtain accurate solutions when using a Dirichlet formulation.  
\begin{figure}[h!]
	\centering
		\includegraphics[width=0.8\textwidth]{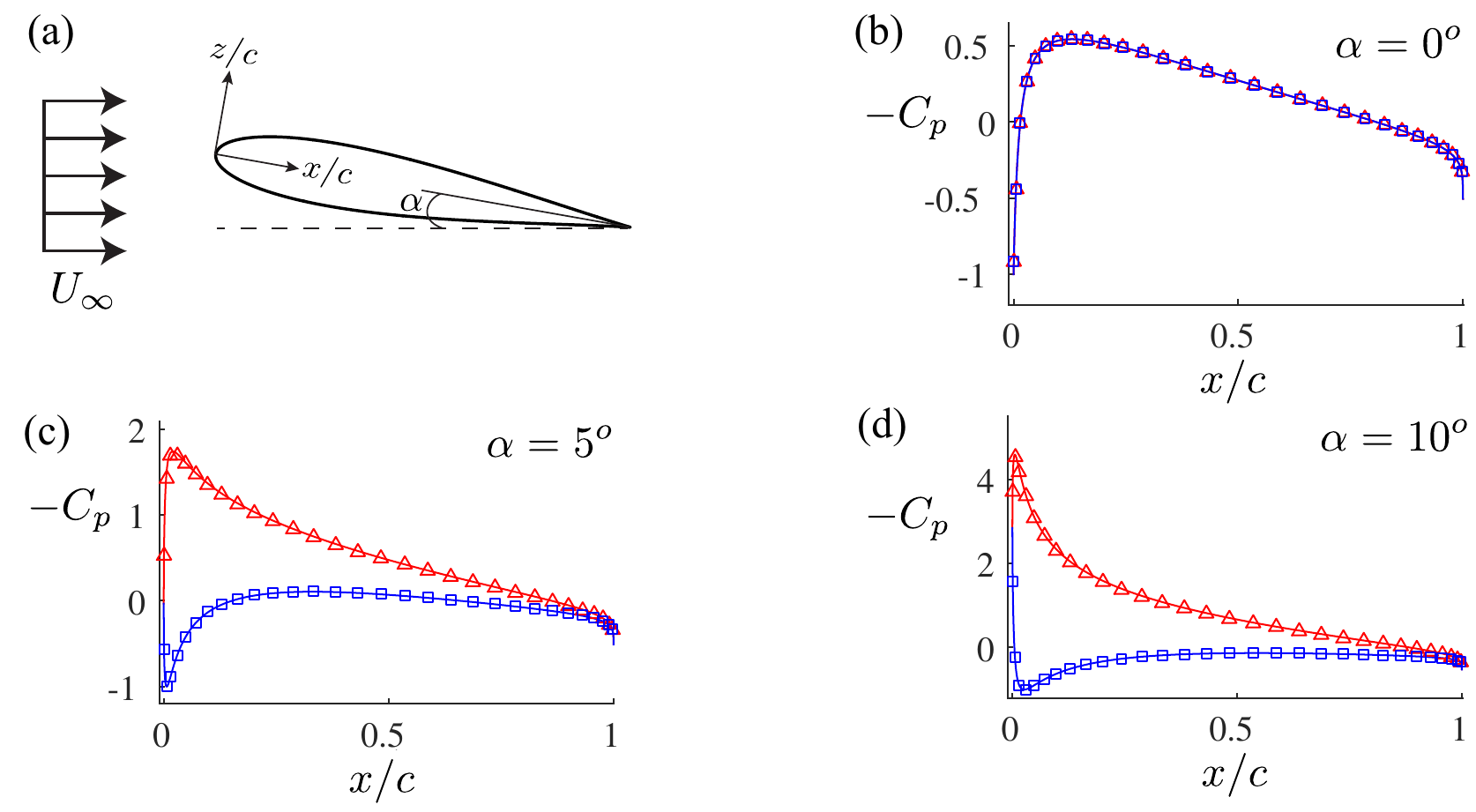}
	\caption{(a) Schematic of a 15\% thick van de Vooren airfoil at an angle of attack, $\alpha$, in a steady flow, $U_\infty$.  The coefficient of pressure over the top and bottom surface of the airfoil for three angles of attack:  (b) $\alpha = 0^0$, (c) $\alpha = 5^0$ and (d) $\alpha = 10^0$.  The analytical solution for the top surface is represented by the solid red line while the numerical solution is represented by the red triangles.  For the bottom surface, the analytical solution is represented by the solid blue line and the numerical solution is represented by the blue squares.}
	\label{fig:CoefPres}
\end{figure}

 The current unsteady three-dimensional formulation is used to solve this two-dimensional steady flow problem by solving for the flow over a van de Vooren wing with a rectangular planform shape and an aspect ratio of $1000$.  The wing is discretized into $30$ chordwise boundary elements for both the top and bottom surfaces and $50$ spanwise boundary elements for a total of $3000$ body elements.  The computation is discretized into $10$ timesteps with the elapsed time for each step being $\Delta t = 100$ s.  The starting vortex present in the unsteady numerical solution is $1000$ chord lengths downstream of the wing since the total elapsed time for the simulations is $\Delta t_{total} = 1000$ s, the free-stream velocity is $U_\infty = 1$ m/s and the chord length of the wing is $c = 1$ m.  This is necessary since the analytical solution assumes that the starting vortex is infinitely far away.  Also, to conform with the analytical solution the wake is `frozen', that is, it is not allowed to roll-up or advect with the local velocity field.  

The coefficient of pressure over the wing is calculated for the section nearest the symmetry plane of the wing where the flow is nearly two-dimensional.  Figure \ref{fig:CoefPres}a shows a schematic of the physical problem where a 15\% thick van de Vooren wing is placed in a steady flow of $U_\infty$ and an angle of attack, $\alpha$.  Figure \ref{fig:CoefPres}b--d compares the exact solutions (solid lines) to the numerical BEM solutions (square and triangle markers).  For $\alpha = 0^o$, the pressure coefficient is observed to stagnate at the leading edge ($C_p = 1$).  Subsequently, the pressure drops as flow is accelerated symmetrically over the top and bottom surfaces until a minimum pressure of $C_p = -0.5$ is reached around $x/c = 0.2$ near the maximum thickness.  After the maximum thickness location there is a gradual pressure recovery of the flow to the trailing edge.  For $\alpha > 0^o$, a prominent leading-edge suction peak can be observed that grows with increasing $\alpha$.  Importantly, there is excellent agreement between the analytical and numerical solutions over a range of angles of attack from $\alpha = 0$ -- $10^o$.  

\subsection{Two-Dimensional Unsteady Flow}
\begin{figure}[h!]
	\centering
		\includegraphics[width=0.8\textwidth]{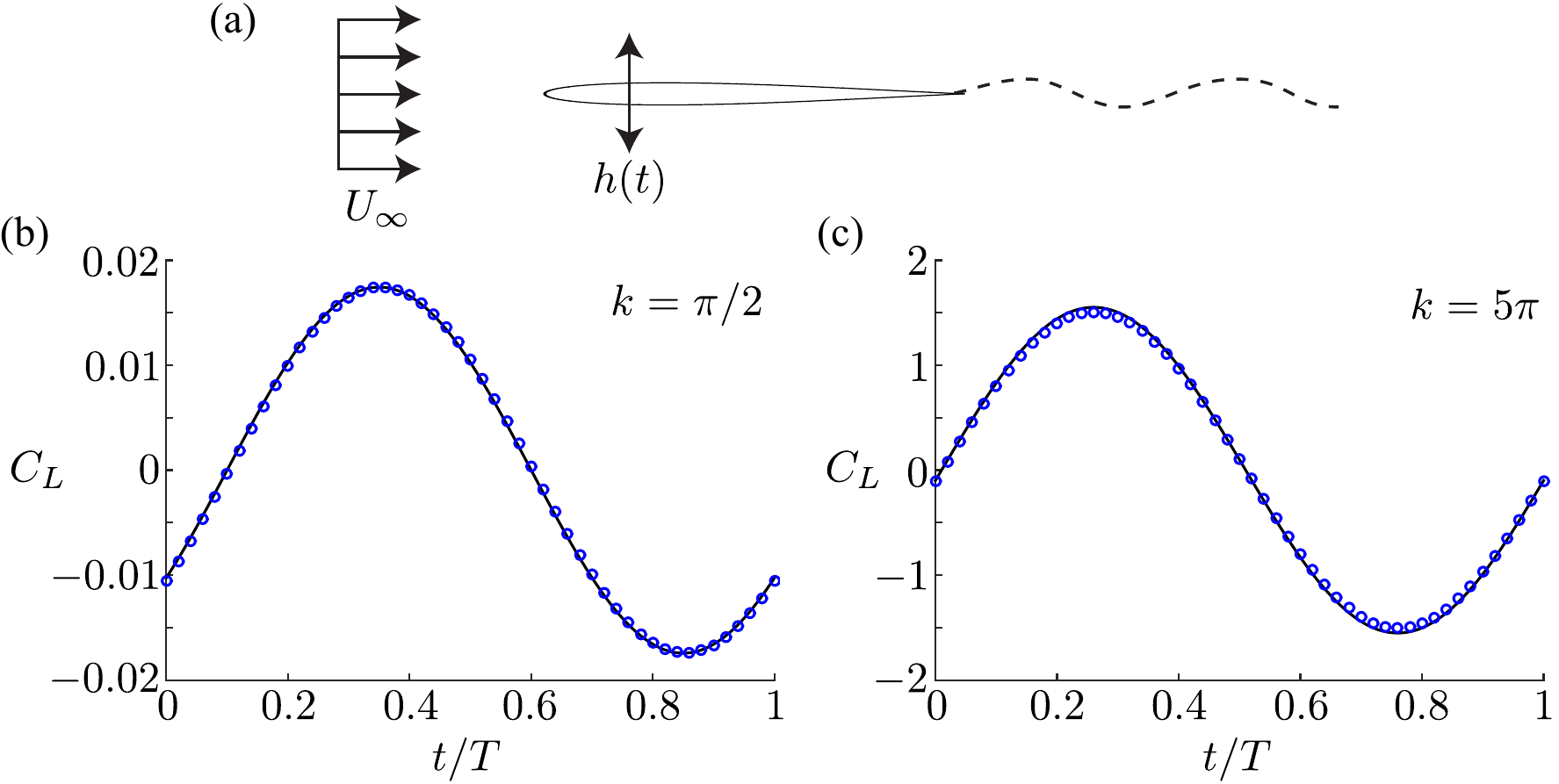}
	\caption{Comparison of the numerical solution marked by circles (o) with the Theodorsen model denoted by the solid line (--).  The parameters of $c = 1$, $U_\infty = 1$, $h/c = 0.001$ were used for both graphs while the left graph used $St = 0.001$ and $k = \pi/2$ and the right graph used $St = 0.01$ and $k = 5 \pi$.}
	\label{Theo}
\end{figure}

The unsteady lift produced by a two-dimensional thin airfoil heaving with small amplitude motion is examined next.  Theodorsen \cite[]{Theodorsen1935} provided an analytical solution to this problem and a schematic detailing the problem is shown in Figure \ref{Theo}a.  The Theodorsen model for a pure heaving motion can be cast into a non-dimensional form,
\begin{align}
 c_l &= -2 \pi^2 \; St \; |C(k)| \; cos(2 \pi \tau + \phi) + \pi^2St\; k \; sin(2 \pi \tau) \\
 & \mbox{where,} \nonumber \\
c_l & = \frac{L}{1/2 \; \rho c U_\infty^2} \quad h = h_0 \; sin (2 \pi \tau) \quad St = \frac{2 h_0 f}{U_\infty} \quad k = \frac{\pi f c}{U_\infty}  \quad \tau = ft \nonumber
\end{align}

\noindent Here, $h$ is the time varying heave, $|C(k)|$ is the magnitude and $\phi$ is the phase angle of the lift deficiency factor, which is a complex number.  A 0.1\% thick symmetric NACA airfoil was used to simulate an infinitesimally thin plate.  The parameters used were $c = 1$ m, $U_\infty = 1$ m/s, $h_0/c = 0.001$ with two cases of $St = 0.001$, $k = \pi/2$ and $St = 0.01$, $k = 5 \pi$.   The three-dimensional formulation is used to solve this two-dimensional unsteady problem by solving for the flow over airfoil with a rectangular planform shape and an aspect ratio of $1000$.  The lift coefficient from the numerical solution is then normalized by the planform area instead of the chord length as in the Theodorsen model.  The wing was discretized into $50$ chordwise boundary elements for both the top and bottom surfaces and $40$ spanwise boundary elements for a total of $4000$ body elements.  The computation was discretized into $50$ timesteps per oscillation cycle for $4$ cycles.  The force calcuations were taken from the fourth oscillation cycle.  A frozen wake was used to simulate the assumptions in the Theodorsen model.  Figure \ref{Theo} shows excellent agreement between the time-varying lift coefficients of the analytical and numerical solutions.  The BEM solution of the time-varying lift coefficient for a Strouhal number of $St = 0.01$ has a slightly lower peak lift coefficient of $C_L = 1.505$ compared to the Theodorsen model of $c_l = 1.545$, which is also seen in other studies \cite[]{Willis2007}.

\subsection{Three-Dimensional Steady Flow}
\begin{figure}[h!]
	\centering
		\includegraphics[width=0.99\textwidth]{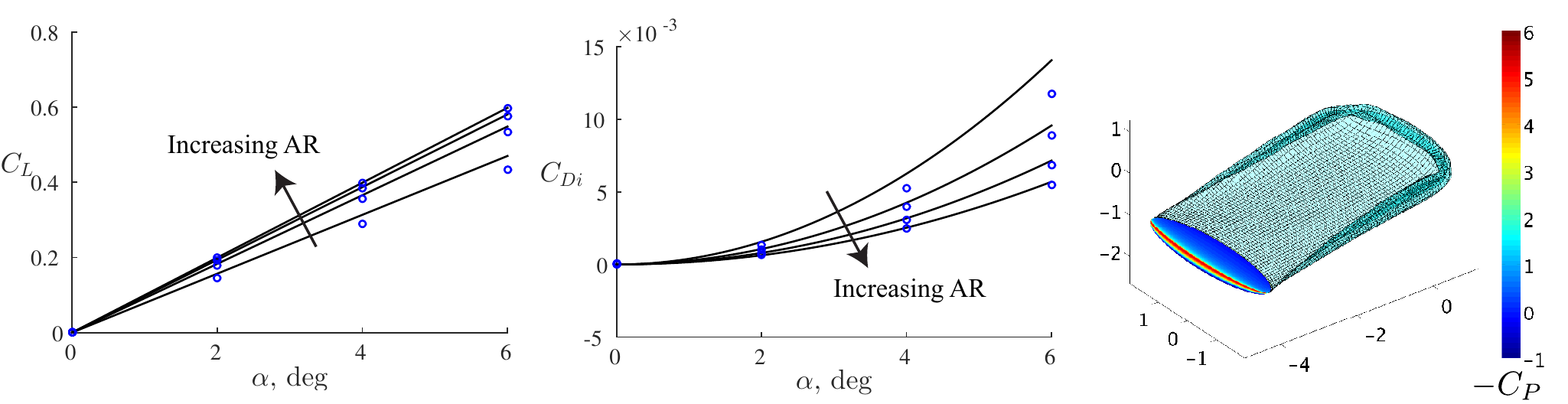}
	\caption{Calculated coefficient of lift and induced drag of steady flow over an elliptical wing at varying angles of attack $\alpha$.  The numerical solution (circles) compare well with the analytical solution (solid line).  There are four aspect ratios used in the calculations: $AR = 5, \: 10, \: 15,$ and $20$.}
	\label{fig:EllpWing}
\end{figure}

The steady flow over a three-dimensional elliptical wing was examined to help validate the three-dimensional aspects of the current formulation.  A teardrop shaped airfoil with a maximum thickness-to-chord ratio of $t/c = 0.02$ was used to simulate a flat plate.  A 2\% thickness was chosen to be sufficiently large so that there is a well-defined leading-edge to avoid numerical sensitivity in the induced drag calculation.  There were 70 chordwise elements on the top and bottom surfaces as well as 50 spanwise elements for a total of $7000$ body elements.  The computation used $4$ timesteps with $\Delta t = 250$ s.  A ``frozen'' wake was used in the force calculations of this validation.  Figure \ref{fig:EllpWing} shows good agreement between the analytical and numerical calculations.  It is expected that the numerical and analytical solutions would deviate at lower aspect ratios.  This is where the finite-wing theory solution begins to breakdown.  An example of the wake structure for a ``free" wake is shown in Figure \ref{fig:EllpWing}c.  In this calculation the wing has an aspect ratio of $AR = 5$, an angle of attack of $\alpha = 15^o$.  The airfoil thickness-to-chord ratio is $t/c = 0.2$.  There are $70$ timesteps with $\Delta t = 0.07$ s between them.  The wake is allowed to rollup showing the characteristic horseshoe vortex system connecting the tip vortex system to the starting vortex.  On the body the colormap represents the pressure coefficient.  There is stagnation pressure near the leading and trailing-edges while there is a minimum in pressure near the leading-edge responsible for leading-edge suction and a majority of the lift production.

\subsection{Three-Dimensional Unsteady Flow}
\begin{figure}[h!]
	\centering
		\includegraphics[width=0.9\textwidth]{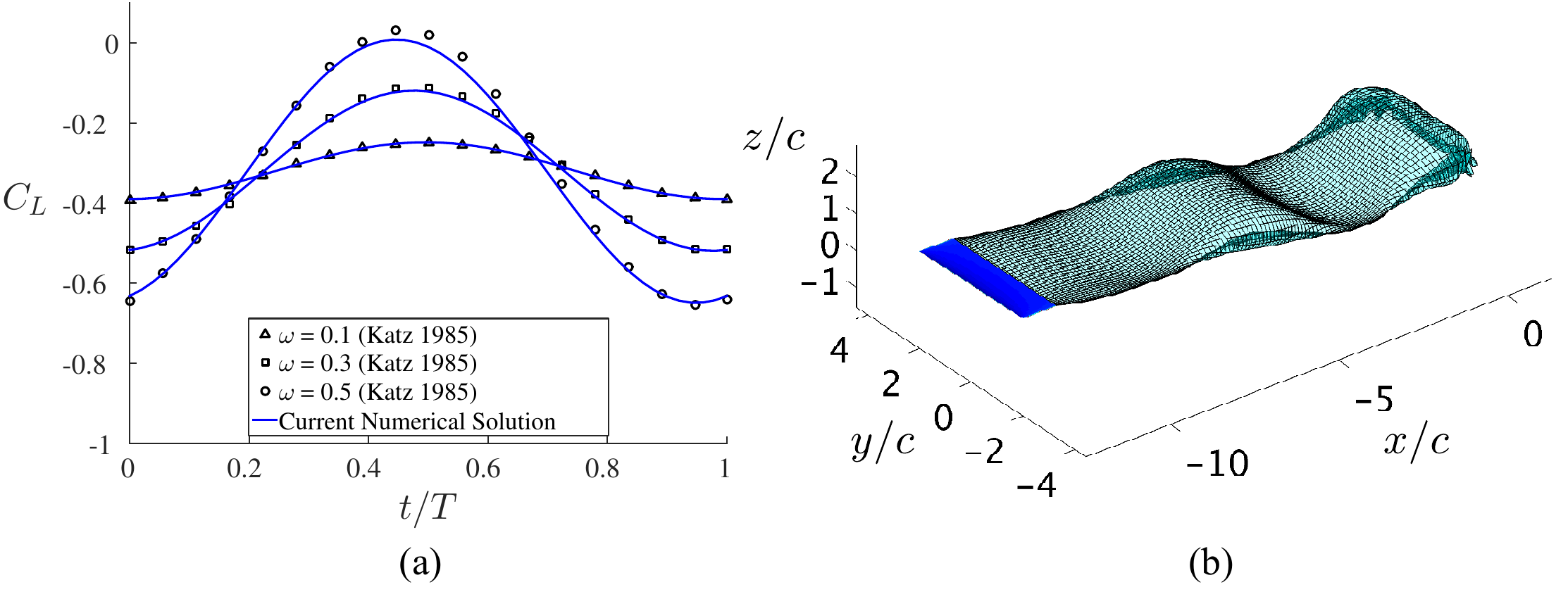}
	\caption{Wing oscillating under unsteady flow conditions.  (a) Lift coefficient calculated with the current numerical method (solid lines) compared to a previous numerical calculation (open markers) by Katz \cite{Katz1985}.  (b)  Boundary element wake deformation.}
	\label{fig:UnsRectWing}
\end{figure}

The unsteady flow problem of the lift on a oscillating three-dimensional wing is examined next.  There is a numerical solution to this problem presented by Katz \cite{Katz1985} where he calculated the lift generated by a heaving three-dimensional rectangular wing at a negative angle of incidence using a boundary element method.  For this case the rectangular wing has an aspect ratio of $AR = 4$, an angle of incidence $\alpha = -5^o$ and a heave amplitude of $h = 0.1c$, where $c$ is the chord length.   The reduced frequency $k = \pi f c/ U_\infty$ is varied for three cases $k = 0.1, \: 0.3,$ and $0.5$ where $f$ is the oscillation frequency and $U_\infty$ is the free-stream velocity.  In the current numerical implementation a $2$\% thick tear-drop shaped airfoil is used to simulate a infinitely-thin flat plate.  The top and bottom surface of the wing is discretized into $30$ chordwise and $50$ spanwise elements for a total of $3000$ body elements.    The computation is discretized into $50$ timesteps per cycle with $\Delta t = [0.63, 0.21, 0.125]$ s for each of the reduced frequency cases respectively.  A total of $4$ oscillating cycles are computed.   Figure \ref{fig:UnsRectWing}a shows the lift coefficient as a function of time over the fourth oscillation cycle.  Excellent agreement can be seen between the current numerical implementation (solid line) and the numerical solution of \cite{Katz1985} (open markers).  As the reduced frequency is increased the magnitude of the unsteady lift is seen to grow and there is a negative phase-shift in the lift oscillation as well.  Figure \ref{fig:UnsRectWing}b shows the rollup of the wake elements that are free to advect with the local velocity field for the case where $k = 0.5$.  As the airfoil heaves upward, the circulation from the unsteady motion adds to the circulation due to the angle of incidence and subsequently the tip vortices increase their strength.  When the airfoil heaves downward the circulation due to the incidence angle and the downward motion counteract each other to produce weak tip vortices.  The net effect is to produce oscillating regions of strong and weak upwash.  

\subsection{Viscous Boundary Layer Solution}
\begin{figure}[h!]
	\centering
		\includegraphics[width=0.9\textwidth]{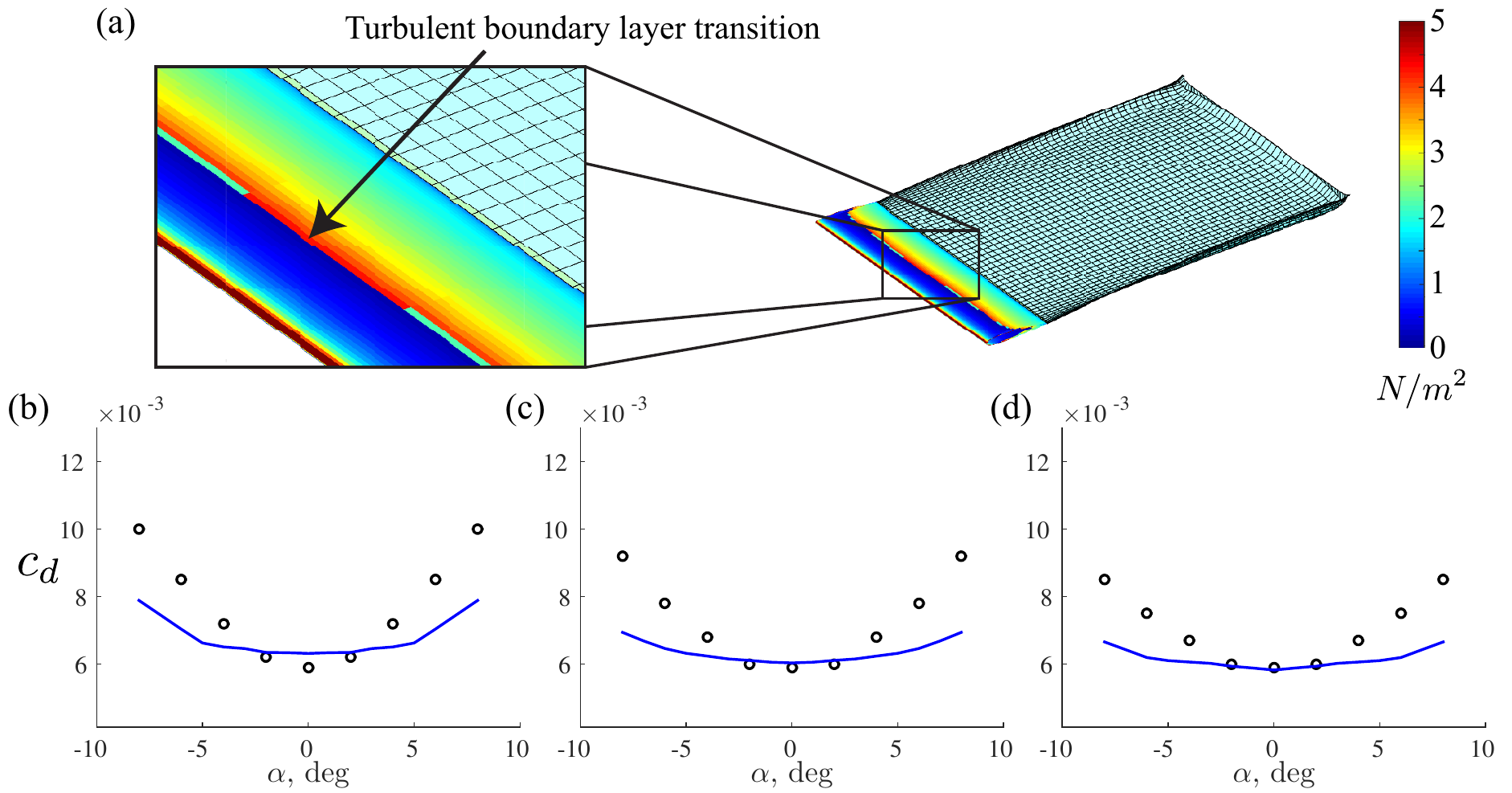}
	\caption{(a) Shear stress on the surface of a NACA 0012 wing in steady flow conditions.  Drag coefficient calculated numerically (solid line) compared to experimental data (circles) from Abbott and von Doenhoff \cite{Abbott1959}: (b) $Re = 3 \times 10^6$, (c) $Re = 6 \times 10^6$, and (d) $Re = 9 \times 10^6$}
	\label{fig:ViscDrag}
\end{figure}

The viscous drag over a NACA 0012 airfoil was examined next by using the viscous boundary layer solver coupled with the outer potential flow solution from the boundary element method.  Figure \ref{fig:ViscDrag}a shows a wing with $AR = 4$ at an angle of attack $\alpha = 3^o$ and a Reynolds number of $Re = 1\times 10^6$.  The boundary elements are colored with the value for the shear stress acting on each element.  When the wing is examined up close, the shear stress is seen to be high at the minimum pressure location on the wing as flow is accelerated around the leading-edge.  Then the shear stress decreases downstream of the minimum pressure point until the boundary layer transitions to a turbulent boundary layer.  At this location, the shear stress is seen to discontinuously increase and then slowly decrease toward the trailing edge.   

A wing with an $AR = 1000$ was used to compare the calculated drag with experimental measurements published in Abbott and von Doenhoff \cite{Abbott1959}.  The calculations spanned a range of angles of attack between $-8^o$ and $8^o$ and three Reynolds numbers of $Re = [3\times 10^6, \; 6\times 10^6, \; 9\times 10^6]$.  The top and bottom surface of the wing is discretized into $50$ chordwise and $50$ spanwise elements for a total of $5000$ body elements.    The computation is discretized into $5$ timesteps with $\Delta t = 200$ s.  The wake is `frozen' in the computations.   

Excellent agreement can be seen between the calculated drag coefficient (solid line) and the experimentally measured values (open circles) for low angles of attack less than $\pm 4^o$.  For higher angles of attack, the drag coefficient is under-predicted by the numerical solution.  In the worst case examined here the calculated drag is $22$\% lower than the measured value at $\alpha = 8^o$.  This discrepancy could be due to mild or moderate trailing-edge separation that may occur in the experiments, but not modeled in the current numerical solution.  Future work of incorporating a separation model may improve the drag calculations.

\subsection{Self-Propelled Biological Propulsion}
\begin{figure}[h!]
	\centering
		\includegraphics[width=0.99\textwidth]{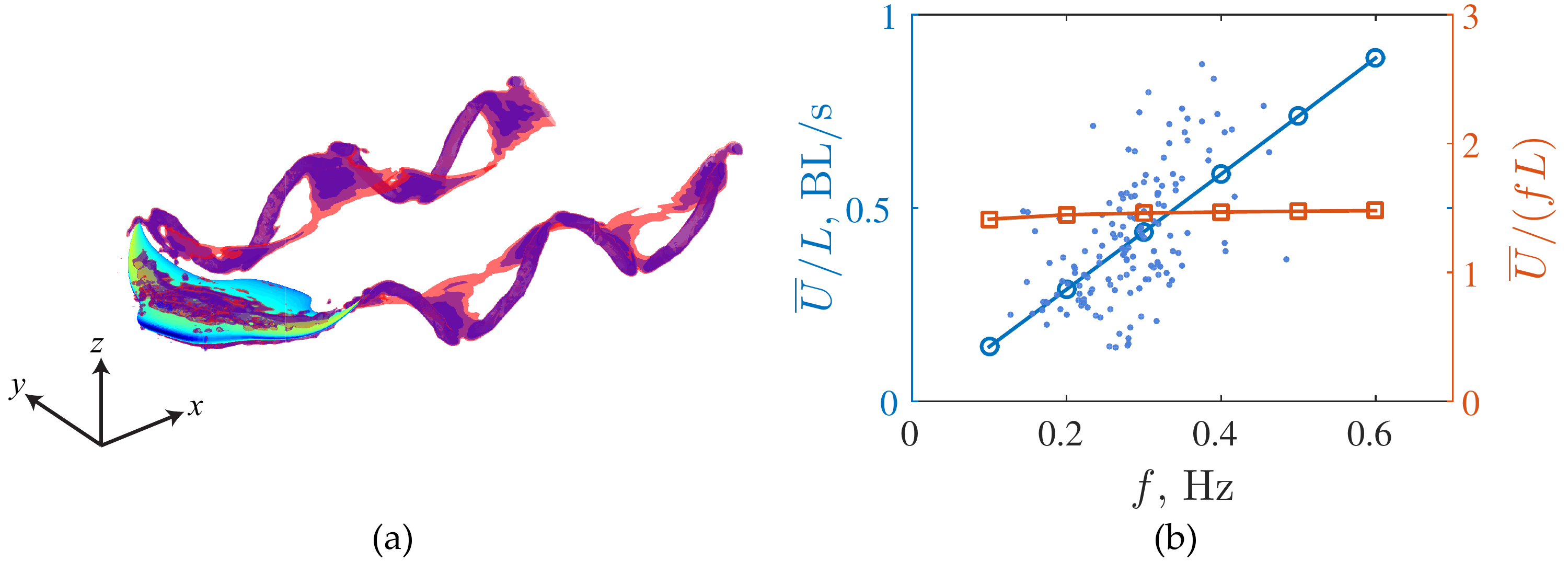}
	\caption{Reproduced from \cite{Fish2016}.  (a) Vortex wake of a manta ray swimmer identified by isosurfaces of the $\lambda_2$ criteria.  (b) Numerical calculations of the self-propelled speed as a function of frequency (solid lines with open markers) compared to biological data (filled markers).  The left vertical axis shows the mean speed normalized by the body length while the right vertical axis shows the mean speed normalized by the frequency and body length.}
	\label{manta}
\end{figure}

Recent work has examined in detail the forces and flow structures exhibited by a numerical manta ray swimmer by using the code presented in the current study \cite{Fish2016}.  In the previous work the numerical method was not shown to be extensively validated as is the case and the focus of the current work.  From extensive biological data gathered in aquaria and in the field, the manta ray swimmer (figure \ref{manta}a) was modeled to be geometrically similar to real mantas and to oscillate its fins with the same kinematic motions of the rays (see \cite{Fish2016} for further details).  The swimming motions of the manta are three-dimensionally complex because of the flexibility of the propulsive pectoral fins. The manta propels itself by vertically moving its enlarged pectoral fins in a flapping motion, combined with waves moving through the fins in both the spanwise and chordwise directions. The motion is thus both oscillatory, and shape-changing (undulatory), although the undulatory component in the chordwise direction is small (the wavelength of the undulation is greater than the chord length of the pectoral fin).  

Figure \ref{manta}a shows the vortex wake produced by a free-swimming manta ray.  The vortices are marked as isosurfaces of the $\lambda_2$ criteria, which distinguishes pressure minima in a plane after discarding unsteady straining and viscous effects \cite{Jeong1995a}.  The manta is seen to shed a series of interlocked vortex rings with one set originating from each pectoral fin.

The left axis line in Figure \ref{manta}b shows the mean swimming speed normalized by the body length as a function of oscillation frequency for the numerical manta ray.  The right axis line further normalizes the mean speed by the oscillation frequency to highlight the near linear dependence upon the oscillation frequency.  The small blue points represent biological data of the swimming speed of real manta rays as a function of their oscillation frequency.  Importantly, the present numerical method shows excellent agreement with the biological data indicating that the current method can predict the performance of real biological self-propelled swimmers.

\section{Results} \label{sec:results}
To investigate the capabilities and limitations of the current BEM implementation in predicting the self-propelled performance and wake structure of a bio-inspired device, the undulatory elliptical fin experiments of Moored \textit{et. al} \cite{Moored2011b} are modeled.  In these experiments a fin having an elliptical planform with an aspect ratio $\AR = 1.6$ and a NACA 0020 cross-section was fabricated.  The root chord length of the fin was $c =0.254$ m. The fin had four actuating spars that were embedded into a PVC polymer and were actuated in such a manner as to produce a chordwise traveling wave (see \cite{Moored2011b} for the experimental details).  The ratio of the traveling-wave wavelength to root chord in the experiments was varied from $\lambda/c = 3-12$.  In the computations, the two cases of $\lambda/c = 6$ and $12$ were modeled where $\lambda/c = 6$ was previously found to be the most efficient case \cite[]{Clark2006,Moored2012}.  The fin had a linearly increasing amplitude from root to tip in the spanwise direction with a peak-to-peak tip amplitude of $A_{pp} = 0.05$ m.  The computational model of the undulatory fin can be seen in Figure \ref{fig:Wake1}a.  The experimental fin was placed into a free-swimming water tank facility where it was attached to a carriage that was supported by low-friction carts on a rail system (Article No.: ME-9454; PASCO scientific).  This setup allowed for unconstrained motion in the streamwise direction.  The root of the fin was located at the free-surface where an acrylic plate was placed to dampen free-surface waves generated by the unsteady undulations of the fin.  

To properly model this experiment several details must be considered.  The free-surface and acrylic plate introduce a no-flux boundary condition that is modeled by using a mirror image fin connected at the root section.  At the same time, the free-surface can produce waves from the streamwise motion of the plate, however, the wave drag from this motion is neglected in the computations.  The acrylic plate though has another drag source, that is, skin friction drag from the production of its boundary layer.  In the computations, the plate geometry is modeled such that the boundary layer solver calculates the skin friction drag from the plate as well as the submerged body.  The forces acting on the mirror image body and mirror image plate are neglected in the free-swimming calculations.  The low-friction carts actually have a finite rolling friction coefficient, which has been measured to be $\mu_r = 0.0065 \pm 0.0002$ \cite[]{Mungan2012}.  The friction coefficient was found to be relatively independent of the speed, that is, it changed by less than 10\% for speeds between $0$--$2$ m/s.  This speed range covers the swimming speeds attained during the experiment, so the friction force resisting the motion, $F_{fr} = \mu_r W$, where $W$ is the net weight of carriage fin system.  This rolling frictional force was included in some of the computations.  The PVC polymer used to fabricate the fin is less dense than water and produces a buoyant force that counteracts the dry weight of the carriage and fin.  The net weight of the whole system when the fin is submerged is estimated at $W \approx 15$ N.

\subsection{Self-Propelled Performance Metrics}
There are several self-propelled performance metrics that are reported in this study.  Time-averaged quantities are marked by an overbar and are always averaged over the last cycle of the computations.  The time-averaged thrust coefficient, net thrust coefficient and power coefficient are defined as
\begin{align}
C_T = \frac{\overline{T}}{1/2 \, \rho \overline{U}^2 S}, \quad \quad C_{T,net} = \frac{\overline{T}_{net}}{1/2 \, \rho \overline{U}^2 S}, \quad \quad C_{P} = \frac{\overline{P}}{1/2 \, \rho \overline{U}^3 S},
\end{align}
 
\noindent respectively.  The time-averaged swimming speed is $\overline{U}$, the time-averaged power input to the fluid is $\overline{P}$, the planform area is $S$ and the fluid density is $\rho$.  The time-averaged thrust force, $\overline{T}$, is the streamwise component of the force from pressure forces alone and the force is considered positive when it is acting on the body in the $-x$ direction.  The time-averaged net thrust force, $\overline{T}_{net}$, is the difference between the thrust force and the total drag force, which includes all of the sources of drag that are present in a given computation, that is, the skin friction drag over the fin and the wave-suppression plate, and the rolling friction drag of the rail-carriage system.

In addition, the Strouhal number, reduced frequency, non-dimensional speed, propulsive efficiency and the swimming economy are reported as
\begin{align}
St = \frac{f A}{\overline{U}}, \quad \quad k = \frac{2 \pi f c}{\overline{U}}, \quad \quad U^* = \frac{\overline{U}}{f c}, \quad \quad \eta = \frac{\overline{T U}}{\overline{P}}, \quad \quad \xi = \frac{\overline{U}}{\overline{P}},
\end{align}

\noindent respectively, where the frequency of oscillation is $f$, the tip amplitude is $A = A_{pp}/2$ and $\overline{T U}$ is the time-averaged useful power output. The non-dimensional speed, $U^*$, represents the distance traveled in chord lengths over a period of oscillation and is effectively the inverse of the reduced frequency.  The propulsive efficiency is the ratio of useful power output to power input, while the swimming economy represents the distance that can be traveled with a unit of energy.  Both energetic swimming metrics are reported since the economy is more readily measured in self-propelled swimming experiments, while the efficiency is non-dimensional and easily comparable between different systems.

\subsection{Discretization Independence}
\begin{figure}[t!]
	\centering
		\includegraphics[width=0.95\textwidth]{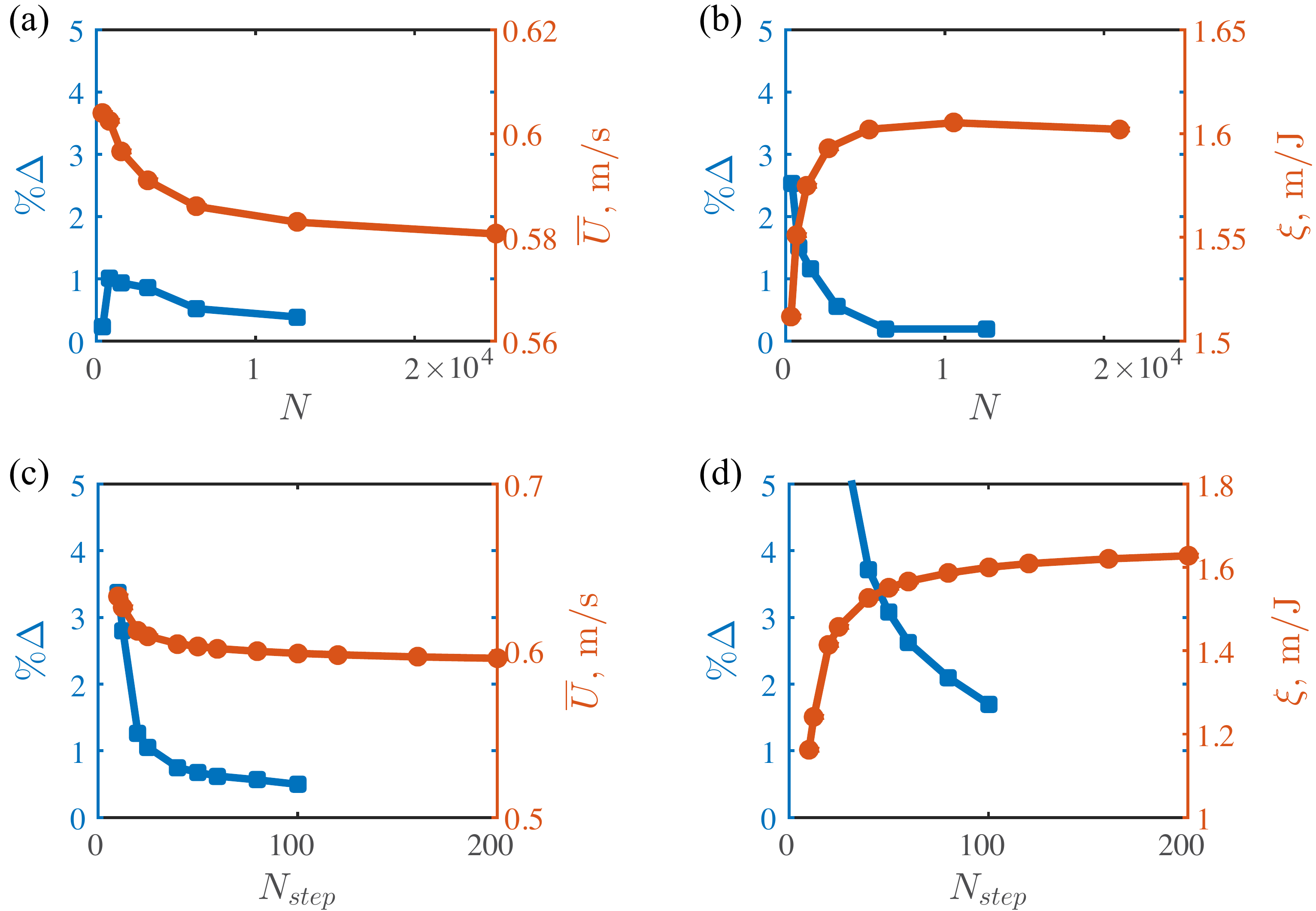}
	\caption{Right axes: Convergence of the time-averaged swimming speed and the swimming economy for varying numbers of the body elements and time steps per oscillation cycle.  Left axes: percent change (\%$\Delta$) in the swimming speed or economy when the number of body elements or time steps per cycle are doubled.  For (a) and (b), the number of time steps per oscillation cycle is fixed at $N_{step} = 50$.  For (c) and (d), the number of body elements is fixed at $N = 800$.}
	\label{fig:converge}
\end{figure}

Convergence of the time-averaged swimming speed, $\overline{U}$, and the swimming economy, $\xi$, was tracked as the number of body elements and the number of time steps per oscillation cycle were varied (Figure \ref{fig:converge}).  The convergence computations are executed for 3 oscillation cycles.  On the right axes in Figure \ref{fig:converge}, good convergence of the swimming speed and swimming economy are shown for both variations in the number of body elements and the number of time steps per oscillation cycle.  The left axes show the percent change (\%$\Delta$) in either the swimming speed or economy when the number of body elements or time steps per cycle are doubled.  It is determined that the free-swimming solution changes by $\mathcal{O}(1\%)$ when $N = 3,200$\footnote{More specifically, there are 40 chordwise elements for the top surface, 40 chordwise elements for the bottom surface and 40 spanwise elements for the combined body of the fin and its image} is doubled and by $\mathcal{O}(2\%)$ when $N_{step} = 80$ is doubled.  These values for $N$ and $N_{step}$ are fixed for all of the following results. 

\subsection{Wake Structure and Self-Propelled Performance}
The self-propelled performance of an undulating elliptical fin is examined in detail to offer a direct comparison of the self-propelled three-dimensional unsteady boundary element implementation formulated in this study with previously published experimental data.  Additionally, the self-propelled performance of other cases that extend beyond the previous experiments are examined to provide novel physical insight into the self-propulsion of three-dimensional bio-inspired fins.  The undulatory fin model examined by the computations more generally represents the fin motions observed in stingrays and pectoral fin swimmers \cite{Rosenberger2001}.  The simulation parameters used for the four numerical cases and the previous experiment data \cite{Moored2011b} are summarized in table \ref{tab:parameters}.  The first numerical case models the exact experimental conditions and parameters used in \cite{Moored2011b}.  Numerical cases 2--4 examine different parameter combinations of amplitude, $A$, and nondimensional wavelength, $\lambda/c$, under no plate drag and no carriage friction conditions.  Plate drag and carriage friction were necessary conditions imposed by the experimental apparatus, however, these conditions are not present in a self-propelled bio-inspired device.  By eliminating these conditions in the numerical simulations, the actual performance of a self-propelled fin can be directly examined.  All of the self-propelled simulations have reached a steady-state condition, that is, the number of oscillation cycles was increased until the cycle-averaged net thrust coefficient was $C_{T,net} = \mathcal{O}(10^{-4})$.   
\begin{table}
   \begin{center}
	\begin{tabular}{c|ccccc} 
  	   &  & \textsc{Plate Drag} & \textsc{Carriage Friction}  & $A$ (m)  & $\lambda/c$  \\  \hline 
	 & & & & & \vspace{-10pt} \\
	Experiments & & Yes & Yes & 0.025 & 6    \\
         Numerical Case 1 & & Yes & Yes & 0.025 & 6    \\
         Numerical Case 2 & & No & No & 0.025 & 6    \\
         Numerical Case 3 & & No & No & 0.025 & 12   \\
	Numerical Case 4 & & No & No & 0.05 & 6   \\
	\end{tabular} 
   \end{center}
  	\caption{Simulation parameters used in the present study and experimental parameters used in \cite{Moored2011b}.}
 	\label{tab:parameters}
\end{table}
\begin{figure}[h!]
	\centering
		\includegraphics[width=0.99\textwidth]{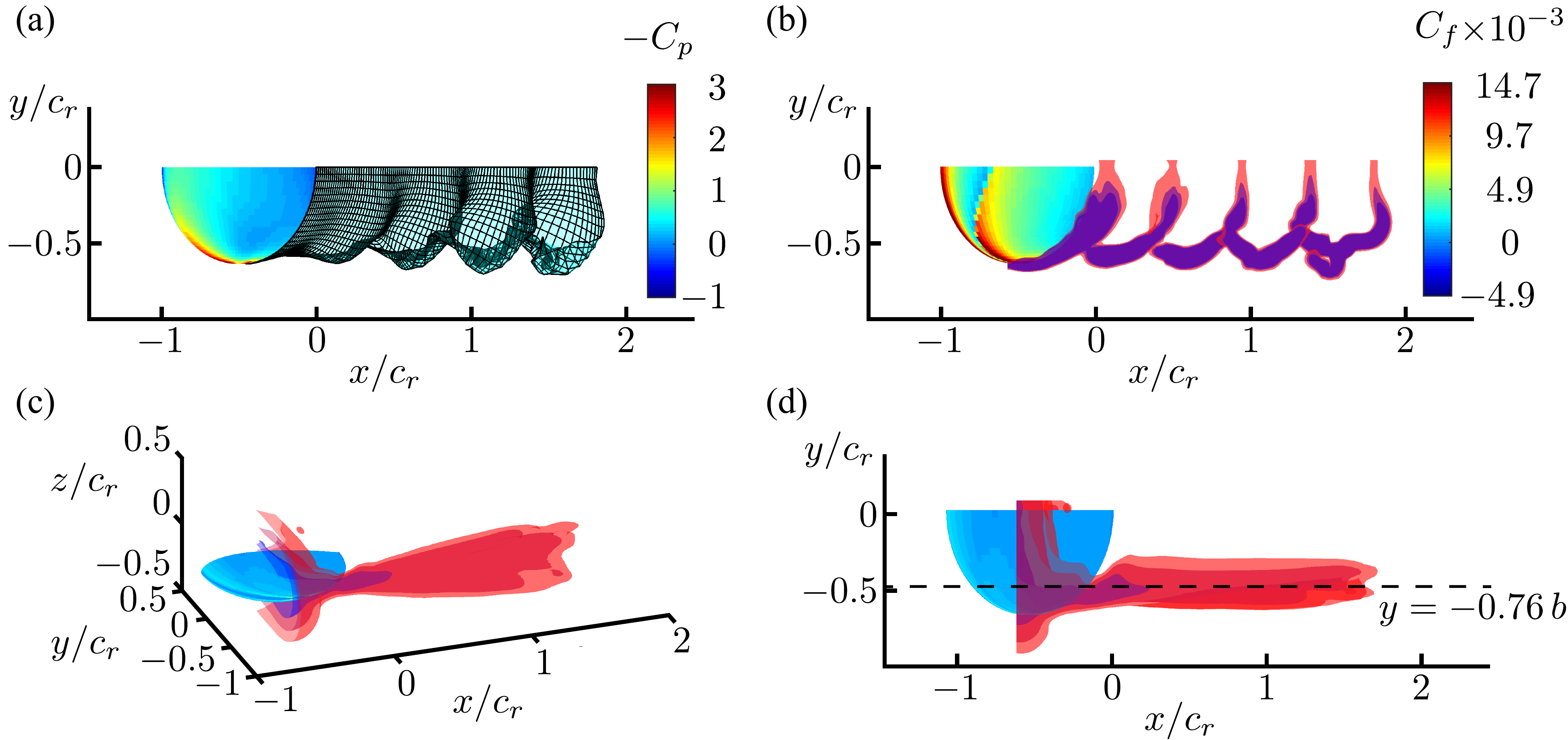}
	\caption{ (a) Boundary element discretization of the wake.  The fin surface colormap represents the pressure coefficient.  (b) Vortex wake identified by isosurfaces of the $\lambda_2$ criteria.  The fin surface colormap represents the skin friction coefficient.  (c)--(d) Isosurfaces of the time-averaged $x$-component of the velocity.  The pink, red and purple isosurfaces are 5\%, 7\% and 10\% above the mean swimming speed.}
	\label{fig:Wake1}
\end{figure}

Figure \ref{fig:Wake1} presents the wake structure of the ellipitcal bio-inspired fin for numerical case 1 (see Table \ref{tab:parameters}).  Figure \ref{fig:Wake1}a shows the wake doublet elements rolling up into coherent vortex structures, which is made evident with the $\lambda_2$ isosurfaces highlighted in  figure \ref{fig:Wake1}b.  It is clear that the elliptical fin is shedding a series of interlocked vortex rings through which fluid is accelerated.  In fact, the time-average of the streamwise velocity is shown for isosurfaces of 5\%, 7\% and 10\% above the free-stream velocity for the pink, red and purple surfaces, respectively (figure \ref{fig:Wake1}c and \ref{fig:Wake1}d).  Note that the isosurface calculation begins at nearly the half-chord of the fin.  The time-averaged jet of fluid accelerated by the vortex rings is observed to have its peak momentum flux at 76\% of the span (Figure \ref{fig:Wake1}d).  This is in good agreement with previous experiments \cite[]{Dewey2011} that found the peak momentum flux to occur at about 80\% of the span.  The surface of the fin is colored with the pressure coefficient in figure \ref{fig:Wake1}a, which is defined as $C_p = P/\, (1/2\, \rho \overline{U}^2$).  The pressure distribution shows a clear leading-edge suction signature occurring near the fin tip that is coincident with the region of the highest momentum flux generated by the fin.  The shear stress distribution computed from the boundary layer solver is shown on the surface of the fin in figure \ref{fig:Wake1}b and is nondimensionalized as a skin friction coefficient, that is $C_f = \tau/ \, (1/2\, \rho \overline{U}^2 )$.  High shear stress can be seen near the leading edge where the flow is accelerated, but the boundary layer remains laminar.  The shear stress rises to high values near the quarter-chord of the fin due to a transition in the boundary layer to turbulence.  This indicates that a majority of the skin friction acting on the fin is coming from a turbulent boundary layer.  All of the other cases show the same salient features of the wake as well as the pressure and shear stress distributions.
\begin{figure}[h!]
	\centering
		\includegraphics[width=0.99\textwidth]{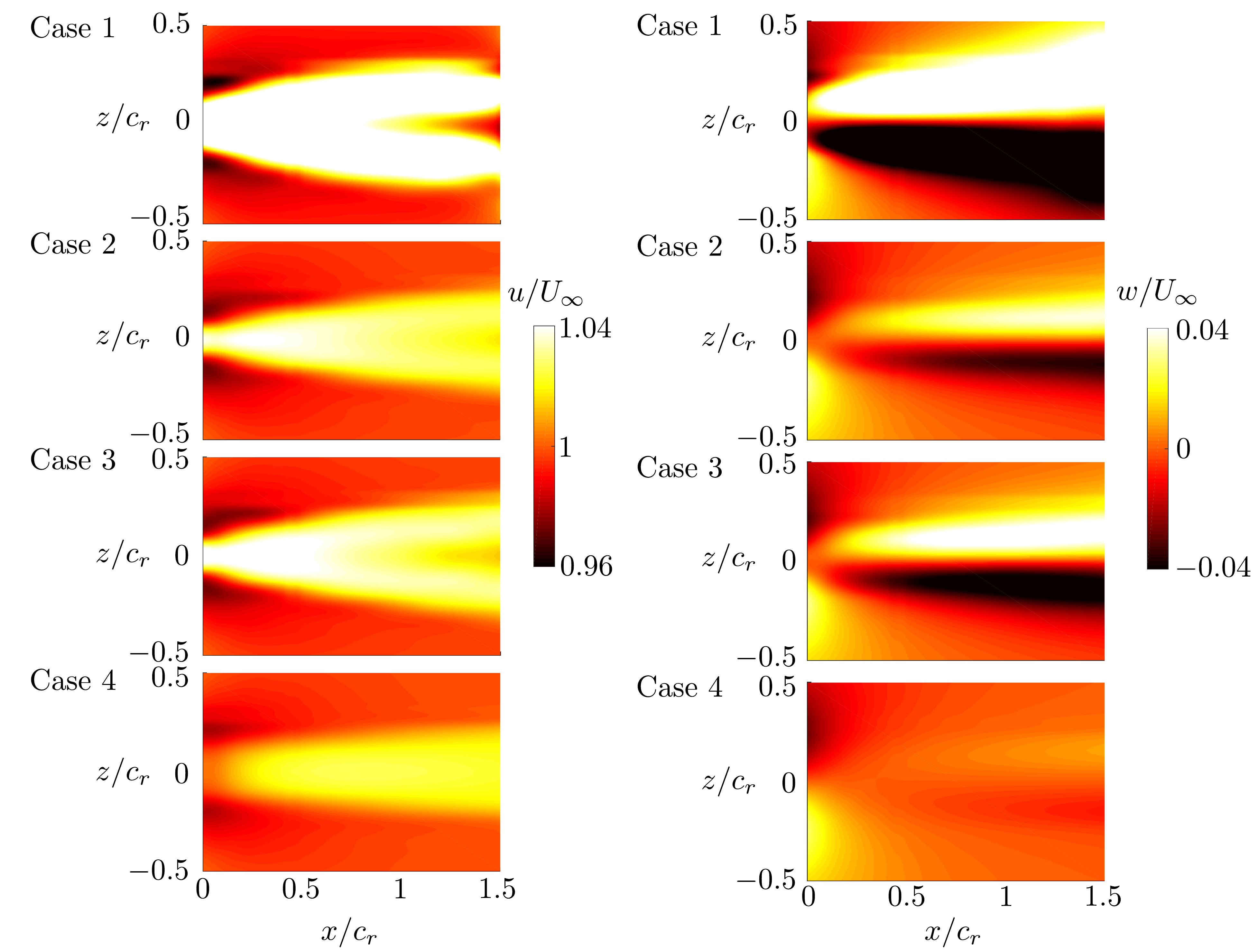}
	\caption{Time-averaged $x$-velocity (left) and $z$-velocity (right) in an $x$-$z$ plane at 76\% of the span for the four numerical cases.}
	\label{fig:Wake2}
\end{figure}

\begin{figure}[h!]
	\centering
		\includegraphics[width=0.99\textwidth]{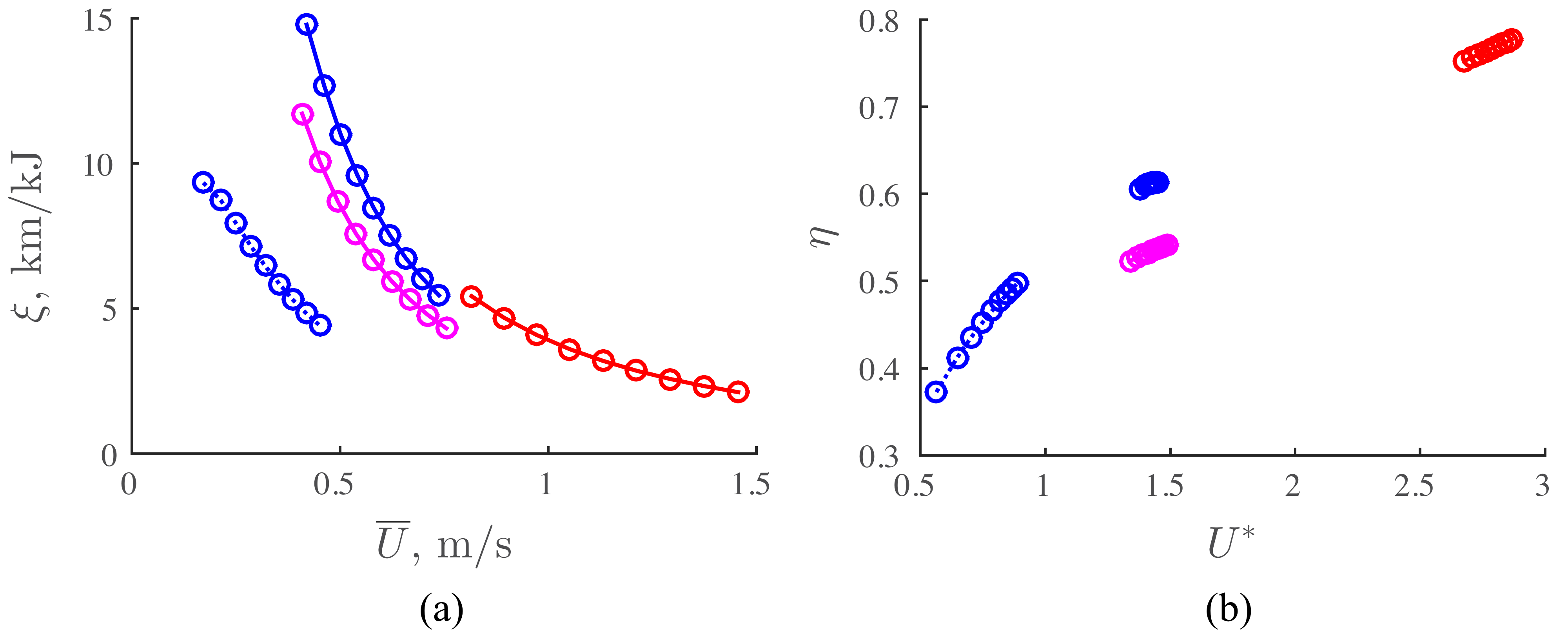}
	\caption{(a) Swimming economy as a function of the self-propelled swimming speed.  (b) Propulsive efficiency as a function of the nondimensional swimming speed.}
	\label{fig:xi_np}
\end{figure}

Although all of the cases produce time-averaged momentum jets, the structure of the momentum jets are quite different.  Figure \ref{fig:Wake2} presents the time-averaged streamwise and cross-stream velocities in the $x$-$z$ plane that cuts through the core of momentum jets at 76\% of the span for all four numerical cases.  The mean streamwise velocity for cases 1--3 show bifurcating jets where the jet splits into two branches.  Bifurcating jets have been implicated in poor efficiency performance \cite[]{Dong2005,Dewey2011}.  In fact, the time-averaged lateral velocity (figure \ref{fig:Wake2} right) highlights that there is excess lateral momentum such that some of the momentum added to the fluid does not perform useful work.  In contrast, case 4 shows a single, non-bifurcating momentum jet and minute amounts of excess lateral momentum.  By considering the magnitude of the lateral momentum relative to the streamwise momentum, the efficiency of each case could be predicted to be ordered from the least efficient to the most efficient as case 1, 3, 2 and 4, respectively.  Indeed, as can be observed later in figure \ref{fig:xi_np} this is exactly the ordering of the cases based on their efficiency.  It is further observed that case 4 has twice the amplitude of motion of the other cases, which leads to the suppression of the jet bifurcation and in turn high efficiency locomotion.

\begin{figure}[h!]
	\centering
		\includegraphics[width=0.99\textwidth]{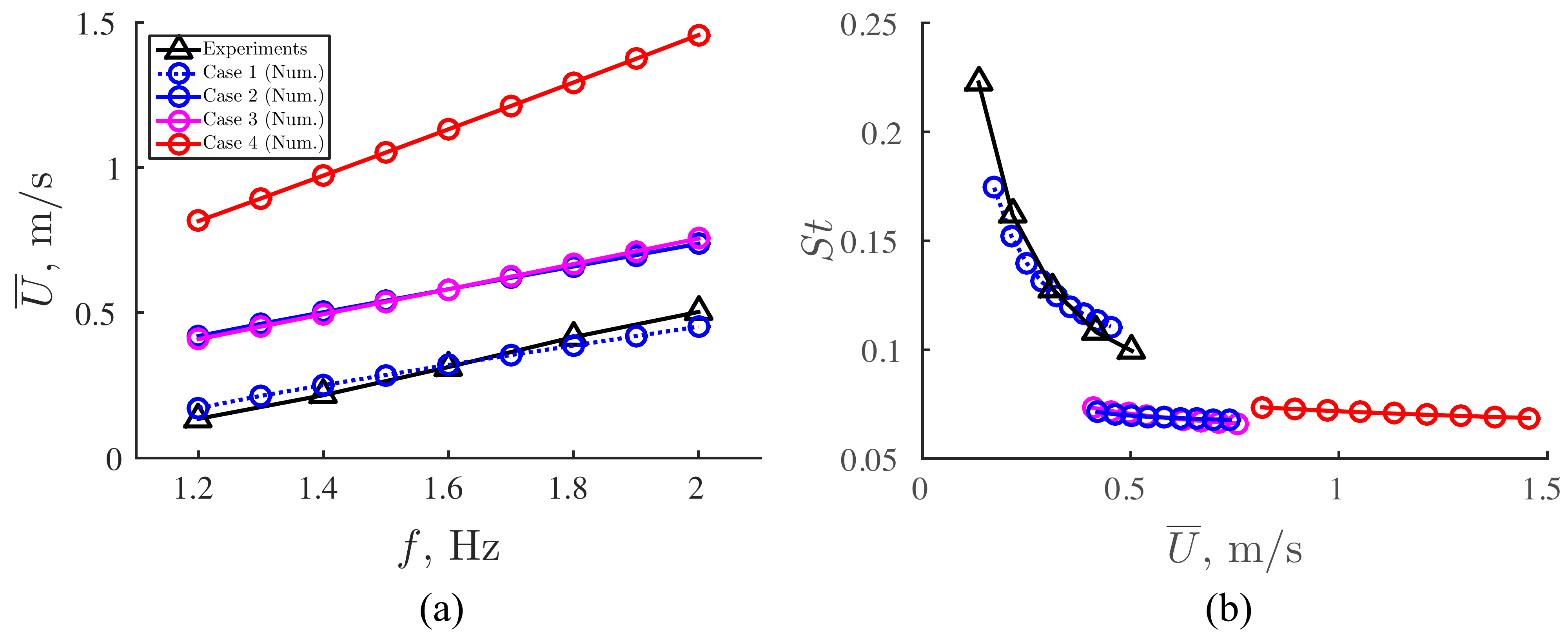}
	\caption{(a) Self-propelled swimming speed as a function of frequency.  (b) Strouhal number as a function of self-propelled swimming speed.  All subsequent experimental and numerical cases marker and line styles follow the legend in (a).}
	\label{fig:U_St_exp}
\end{figure}

Figure \ref{fig:U_St_exp}a presents the self-propelled swimming speed of the elliptical fin for a range of frequencies of $f = 1.2-2$ Hz.  The experimental data \cite{Moored2011b} shows that the swimming speed of the undulating fin increases nearly linearly with an increasing frequency of motion.  Similarly, the numerical simulations show a nearly linearly increasing swimming speed with increasing frequency, however, with a slightly lower slope.  Regardless of this discrepancy, there is still excellent agreement between the numerical simulations and the experimental data further validating the self-propelled numerical implementation presented in this study.  It is expected that viscous effects in the experiments such as small regions of separated flow near the trailing-edge, the three-dimensional boundary layers, or boundary layer thinning from the unsteady motions \cite{Ehrenstein2013,Ehrenstein} can account for the differences among these data.

As expected, when the experimental conditions of plate drag and carriage friction are eliminated (cases 2--4) the undulating fin is observed to have a higher self-propelled swimming speed than case 1 at the same frequency of motion (Figure \ref{fig:U_St_exp}a).  When the nondimensional wavelength of motion is doubled from $\lambda/c = 6$ to $\lambda/c = 12$ there is a slight increase in the slope of the swimming speed curve, but otherwise a negligible effect on the swimming speed.  However, when the amplitude of motion is doubled from $A = 0.025$ m to $A = 0.05$ m, the swimming speed is nearly doubled for all frequencies examined (percent increase of 93--97\%).

Figure \ref{fig:U_St_exp}b presents the Strouhal number as a function of the swimming speed.  Again, the numerical data (case 1) is observed to show excellent agreement with the experimentally measured Strouhal number and both curves show decreasing $St$ with an increase in the frequency of motion and in turn the swimming speed.  Cases 2--4 exhibit a nearly constant $St$, however, they show a slightly decreasing trend with an increase in the swimming speed.  In fact for cases 2--4 the swimming speed does indeed increase proportionally with the frequency leading to a nearly constant $St$.  In contrast, the additional friction from the carriage and additional drag from the plate modify the swimming speed to be a nonlinear function of the frequency that is not obvious in figure \ref{fig:U_St_exp}a, but is made evident by the nonlinear relationship between the $St$ and the swimming speed.

Figure \ref{fig:xi_np}a presents the swimming economy as a function of the swimming speed.  For all cases, as the swimming speed increases there is an inverse relationship with the economy, that is, when swimming faster the distance that can be travelled for a unit of energy is smaller.  This trend has been observed in the self-propelled swimming of manta rays \cite{Fish2016} and in the self-propelled swimming of heaving flexible panels \cite{Quinn2014b}.  However, this trend contradicts the measurements of swimming economy observed in \cite{Moored2011b} and indicates that there was likely an error in those power measurements given the excellent agreement in swimming speed between the numerics and experiments.  Intuitively, one would expect the economy to, in general, have an inverse trend with swimming speed as observed in the current study since the reverse would imply that it would take little energy to swim a given distance for the fastest swimming speeds, which seems unlikely.  To support this intuition, when the economy relationship is examined, 
\begin{align}
\xi = \frac{\overline{U}}{\overline{P}} = \frac{1}{1/2 \, \rho \overline{U}^2 S} \left(\frac{1}{C_P}\right),
\end{align} 
\noindent it can be observed that the economy has an inverse relationship with the swimming speed and the power coefficient, which is also a function of the swimming speed.  Furthermore, experimental power measurements have been reported in \cite{Clark2006} for the same undulating fin as in \cite{Moored2011b}, except that the fin was held in a water channel at a fixed velocity instead of operating in a self-propelled swimming state.  If these power measurements are comparable, then \cite{Clark2006} indicates a 5 fold drop in power when the Strouhal number drops from $St = 0.225$ to $St = 0.1$, precisely the $St$ range observed in \cite{Moored2011b} that in turn corresponds to the speed range of $0.134 \leq \overline{U} \leq 0.503$.  Now, the highest speed, $\overline{U}_H$, in \cite{Moored2011b} relates to the lowest speed, $\overline{U}_L$, as $\overline{U}_H = 3.75\, \overline{U}_L$ and the subsequent power coefficients are related by $C_{P,H} = 1/5\, C_{P,L}$.  By forming a ratio of the economy at the highest swimming speed with the economy at the lowest swimming speed the trend in the economy can be estimated as,  
\begin{align}
\mathcal{R} = \frac{\xi_H}{\xi_L} = \frac{\overline{U}_L^2 C_{P,L}}{\overline{U}_H^2 C_{P,H}} = \frac{5}{3.75^2} = 0.3556
\end{align}

\noindent When the ratio $\mathcal{R} < 1$ the economy is estimated as having an inverse trend with swimming speed and when the $\mathcal{R} > 1$ \textit{vice versa}.  The ratio based on the previous power measurements from \cite{Clark2006} is $\mathcal{R} = 0.3556$ estimating that the economy should drop by nearly one-third in the experiments over the tested speed range.  This estimate further supports the findings of the current numerical study and the idea that the previous experiments have an error in their power measurements.  In fact, the ratio of economies using the numerical data (case 1) from figure \ref{fig:xi_np} is $\mathcal{R} = 0.4725$, which is in good agreement with the expected ratio of economies from the experiments.  

Figure \ref{fig:xi_np}b presents the propulsive efficiency as a function of the non-dimensional swimming speed.  Even though the economy is low for the high swimming speeds and for the high amplitude case (case 4), the efficiency is the highest for these cases with a peak efficiency for case 4 of $\eta \approx 78\%$.  Also, the peak efficiency for case 1 is $\eta \approx 50\%$, which is in good agreement with the experimentally measured peak efficiency being $\eta \approx 55\%$ \cite{Clark2006}.  Interestingly, if the experiments were fully self-propelled without the additional plate drag and carriage friction the numerical results indicate that a peak efficiency of $\eta \approx 61\%$ could be attained.  Now, it is also clear that even though the higher nondimensional wavelength ($\lambda/c = 12$, case 3) did not produce significantly different swimming speeds than case 2, it did reduce the propulsive efficiency to a peak value of $\eta \approx 54\%$.
\begin{figure}[h!]
	\centering
		\includegraphics[width=0.99\textwidth]{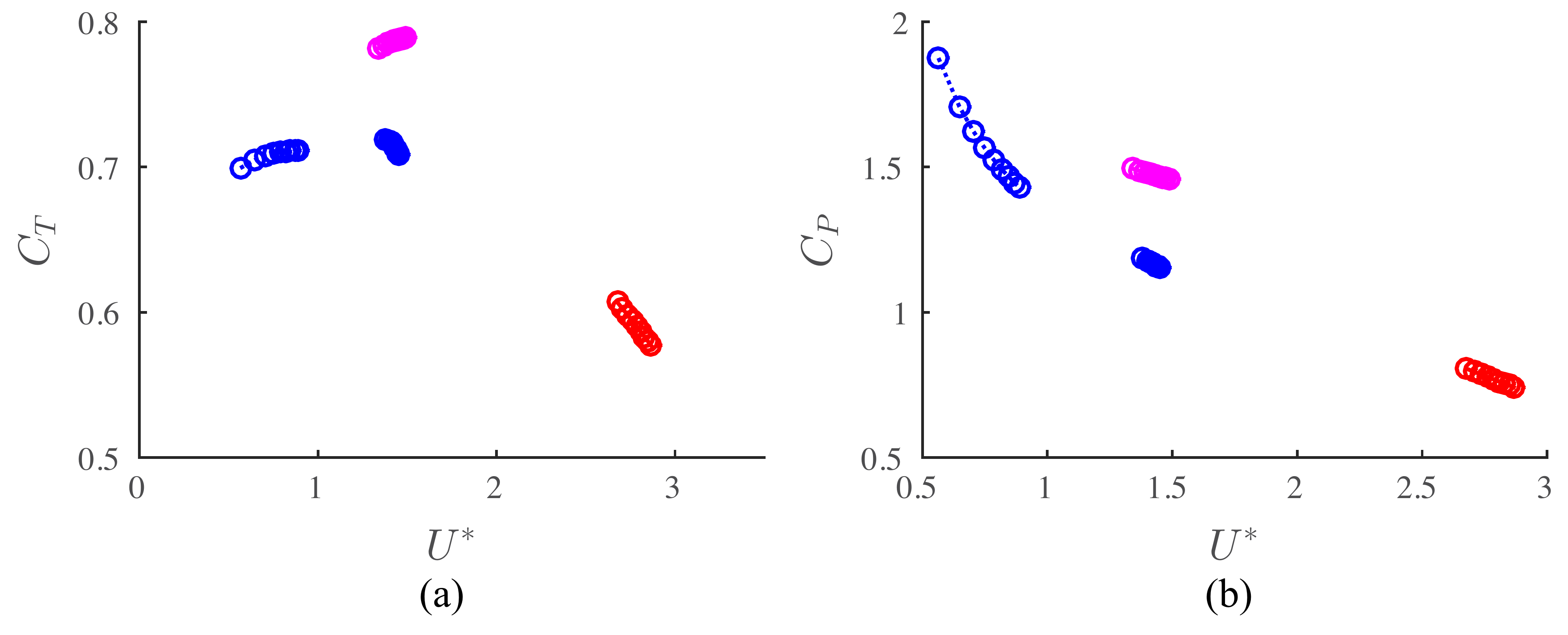}
	\caption{(a) Thrust coefficient as a function of the nondimensional swimming speed.  (b) Power coefficient as a function of the nondimensional swimming speed.}
	\label{fig:Ct_Cp}
\end{figure}

The trends in propulsive efficiency can be further decomposed by examining the thrust (figure \ref{fig:Ct_Cp}a) and power coefficients (figure \ref{fig:Ct_Cp}b) since $\eta = C_T/C_P$.  It is observed that the thrust coefficient for $\lambda/c = 12$ (case 3) is actually higher than $\lambda/c = 6$ (case 2) indicating that the swimming speed is not linearly proportional to changes in $C_T$, but instead is proportional to the frequency and amplitude of motion as proposed in \cite{Bainbridge1957}.  It is clear though that the power coefficient rise is greater than the thrust coefficient rise leading to a lower efficiency.  Additionally, the higher amplitude motion (case 4) is observed to slightly reduce the thrust coefficient as compared to case 2.  At the same time the power is also significantly reduced leading to higher efficiency.  Case 2 shows both a higher thrust coefficient and a lower power coefficient than case 1 leading to an increase the propulsive efficiency.

\section{Conclusions}
A novel boundary element method implementation is developed and presented to examine problems of biological and bio-inspired self-propelled locomotion.  The formulation uniquely combines an unsteady three-dimensional boundary element fluid solver, a boundary layer solver and an equation of motion solver.  The method is validated through a series of analytical, numerical and experimental data that include comparisons with two-dimensional and three-dimensional steady flow solutions, two-dimensional and three-dimensional unsteady flow solutions, viscous drag measurements, self-propelled biological field measurements and self-propelled bio-inspired laboratory measurements.  The method is then employed to extend previous experimental results on a ray-inspired model of a self-propelled undulating fin.  It is discovered that high propulsive efficiencies of 78\% can be obtained when large amplitude motion is used.  Additionally, these motions are shown to suppress the formation of a time-averaged bifurcating jet and instead form a single core jet with negligible time-averaged lateral velocity.  This maximizes the amount of streamwise momentum compared to the amount of wasted lateral momentum and in turn leads to high propulsive efficiency.

\subsection*{Acknowledgments}

This work was supported by the Office of Naval Research under Program Director Dr. Robert Brizzolara,  MURI grant number N00014-08-1-0642.  I would like to acknowledge the helpful discussions with Dan Quinn and Peter Dewey on the design of the experimental undulating elliptical fin.

\bibliographystyle{plain}
\bibliography{BEM_Validation_Literature}

\begin{thebibliography}{10}

\bibitem{Abbott1959}
I.~H. Abbott and A.~E. von Doenhoff.
\newblock {\em {Theory of wing sections}}.
\newblock Dover Publications, 1959.

\bibitem{Bainbridge1957}
B.~Y.~R. Bainbridge.
\newblock {The speed of swimming of fish as related to size and to the
  frequency and amplitude of the tail beat}.
\newblock {\em Journal of Experimental Biology}, 35(1937):109--133, 1957.

\bibitem{Basu1978}
B.~C. Basu and G.~J. Hancock.
\newblock {The unsteady motion of a two-dimensional aerofoil in incompressible
  inviscid flow}.
\newblock {\em Journal of Fluid Mechanics}, 87(1):159--178, 1978.

\bibitem{Borazjani2008}
I.~Borazjani, L.~Ge, and F.~Sotiropoulos.
\newblock {Curvilinear immersed boundary method for simulating fluid structure
  interaction with complex 3D rigid bodies}.
\newblock {\em Journal of Computational Physics}, 227(16):7587--7620, 2008.

\bibitem{Cheng2001}
J.~Cheng and G.~L. Chahine.
\newblock {Computational hydrodynamics of animal swimming: boundary element
  method and three-dimensional vortex wake structure}.
\newblock {\em Comparative Biochemistry and Physiology - A Physiology},
  131:51--60, 2001.

\bibitem{Clark2006}
R.~P. Clark and a.~J. Smits.
\newblock {Thrust production and wake structure of a batoid-inspired
  oscillating fin}.
\newblock {\em Journal of Fluid Mechanics}, 562:415--429, 2006.

\bibitem{Dewey2011}
P.~A. Dewey, A.~Carriou, and A.~J. Smits.
\newblock {On the relationship between efficiency and wake structure of a
  batoid-inspired oscillating fin}.
\newblock {\em Journal of Fluid Mechanics}, 691:245--266, 2011.

\bibitem{Dong2005}
H.~Dong, R.~Mittal, M.~Bozkurttas, and F.~Najjar.
\newblock {Wake Structure and Performance of Finite Aspect-Ratio Flapping
  Foils}.
\newblock {\em 43rd AIAA Aerospace Sciences Meeting and Exhibit}, 2005.

\bibitem{Ehrenstein2013}
U.~Ehrenstein and C.~Eloy.
\newblock {Skin friction on a moving wall and its implications for swimming
  animals}.
\newblock {\em Journal of Fluid Mechanics}, 718:321--346, 2013.

\bibitem{Ehrenstein}
U.~Ehrenstein, M.~Marquillie, and C.~Eloy.
\newblock {Skin friction on a flapping plate in uniform flow}.
\newblock {\em Proceedings of the Royal Society A: Mathematical, Physical and
  Engineering Sciences}, 372(2020):20130345, 2014.

\bibitem{Fish2016}
F.~E. Fish, C.~M. Schreiber, K.~W. Moored, G.~Liu, H.~Dong, and H.~Bart-Smith.
\newblock {Hydrodynamic performance of aquatic flapping: efficiency of
  underwater flight in the manta}.
\newblock {\em Aerospace}, 3(20):1--30, 2016.

\bibitem{Haberman2004}
R.~Haberman.
\newblock {\em {Applied partial differential equations: with Fourier series and
  boundary value problems}}.
\newblock Pearson Prentice Hall, Upper Saddle River, NJ, 4th edition, 2004.

\bibitem{Hess1972}
J.~L. Hess.
\newblock {Calculation of potential flow about arbitrary three-dimensional
  lifting bodies}.
\newblock Technical report, Douglas Aircraft Company, Long Beach, California,
  1972.

\bibitem{Jeong1995a}
J.~Jeong and F.~Hussain.
\newblock {On the identification of a vortex}.
\newblock {\em Journal of Fluid Mechanics}, 285:69--94, 1995.

\bibitem{Kagami1984}
S.~Kagami and I.~Fukai.
\newblock {Application of boundary-element method to electromagnetic field
  problems}.
\newblock {\em Microwave Theory and Techniques, IEEE Transactions on},
  32(4):455--461, 1984.

\bibitem{Katz1985}
J~Katz.
\newblock {Calculation of the aerodynamic forces on automotive lifting
  surfaces}.
\newblock 107:438--443, 1985.

\bibitem{Katz2001}
J.~Katz and A.~Plotkin.
\newblock {\em {Low-speed aerodynamics}}.
\newblock Cambridge University Press, New York, NY, second edition, 2001.

\bibitem{Krasny1986}
R.~Krasny.
\newblock {Desingularization of Periodic Vortex Sheet Roll-up}.
\newblock {\em Journal of Computational Physics}, 65:292--313, 1986.

\bibitem{Liu1997}
P.~Liu and N.~Bose.
\newblock {Propulsive performance from oscillating propulsors with spanwise
  flexibility}.
\newblock {\em Proceedings of the Royal Society A: Mathematical, Physical and
  Engineering Sciences}, 453:1763--1770, 1997.

\bibitem{Liu1999}
P.~Liu and N.~Bose.
\newblock {Hydrodynamic characteristics of a lunate shape oscillating
  propulsor}.
\newblock {\em Ocean Engineering}, 26:519--529, 1999.

\bibitem{Maskew1987}
B.~Maskew.
\newblock {Program VSAERO theory document}.
\newblock Technical report, NASA CR-4023, 1987.

\bibitem{Moored2011b}
K.~W. Moored, P.~A. Dewey, M.~C. Leftwich, H.~Bart-Smith, and A.~J. Smits.
\newblock {Bioinspired propulsion mechanisms based on manta ray locomotion}.
\newblock {\em Marine Technology Society Journal}, 45(4):110--118, 2011.

\bibitem{Moored2012}
K.~W. Moored, P.~A. Dewey, A.~J. Smits, and H.~Haj-Hariri.
\newblock {Hydrodynamic wake resonance as an underlying principle of efficient
  unsteady propulsion}.
\newblock {\em Journal of Fluid Mechanics}, 708:329--348, 2012.

\bibitem{Mungan2012}
C.~E. Mungan.
\newblock {Rolling friction on a wheeled laboratory cart}.
\newblock {\em Physics Education}, 47(3):288--292, 2012.

\bibitem{Pan2012}
Y.~Pan, X.~Dong, Q.~Zhu, and D.~K.~P. Yue.
\newblock {Boundary-element method for the prediction of performance of
  flapping foils with leading-edge separation}.
\newblock {\em Journal of Fluid Mechanics}, 698:446--467, 2012.

\bibitem{Portela1992}
A.~Portela, M.~H. Aliabadi, and D.~P. Rooke.
\newblock {The dual boundary element method: Effective implementation for crack
  problems}.
\newblock {\em International Journal for Numerical Methods in Engineering},
  33:1269--1287, 1992.

\bibitem{Pozrikidis2001}
C.~Pozrikidis.
\newblock {Interfacial dynamics for Stokes flow}.
\newblock {\em Journal of Computational Physics}, 169(2):250--301, 2001.

\bibitem{Quinn2014b}
D.~B. Quinn, G.~V. Lauder, and A.~J. Smits.
\newblock {Scaling the propulsive performance of heaving flexible panels}.
\newblock {\em Journal of Fluid Mechanics}, 738:250--267, 2014.

\bibitem{Quinn2014}
D.~B. Quinn, K.~W. Moored, P.~A. Dewey, and A.~J. Smits.
\newblock {Unsteady propulsion near a solid boundary}.
\newblock {\em Journal of Fluid Mechanics}, 742:152--170, 2014.

\bibitem{Robinson1988}
D.~E. Robinson.
\newblock {\em {Implementation of a separated flow panel method for wall
  effects on finite swept wings}}.
\newblock Phd, Massachusetts Institute of Technology, 1988.

\bibitem{Rosenberger2001}
L.~J. Rosenberger.
\newblock {Pectoral fin locomotion in batoid fishes: undulation versus
  oscillation.}
\newblock {\em The Journal of Experimental Biology}, 204:379--394, 2001.

\bibitem{Shoele2010}
K.~Shoele and Q.~Zhu.
\newblock {Numerical simulation of a pectoral fin during labriform swimming}.
\newblock {\em The Journal of Experimental Biology}, 213:2038--2047, 2010.

\bibitem{Shoele2009}
Kourosh Shoele and Qiang Zhu.
\newblock {Fluid-structure interactions of skeleton-reinforced fins:
  performance analysis of a paired fin in lift-based propulsion.}
\newblock {\em The Journal of experimental biology}, 212:2679--2690, aug 2009.

\bibitem{Smith1996}
M.~J.~C. Smith.
\newblock {Simulating moth wing aerodynamics: Towards the development of
  flapping-wing technology}.
\newblock {\em AIAA Journal}, 34(7):1348--1355, 1996.

\bibitem{Theodorsen1935}
T.~Theodorsen.
\newblock {General theory of aerodynamic instability and the mechanism of
  flutter}.
\newblock Technical report, NACA report No. 496, 1935.

\bibitem{VandeVooren1969}
A.~I. van~de Vooren and L.~S. de~Jong.
\newblock {Calculation of incompressible flow about aerofoils using source,
  vortex and doublet distributions}.
\newblock Technical report, TW-86 of the Math. Inst. of the University of
  Groningen, The Netherlands, 1969.

\bibitem{Voutsinas2006}
S.~G. Voutsinas.
\newblock {Vortex methods in aeronautics: how to make things work}.
\newblock {\em International Journal of Computational Fluid Dynamics},
  20(1):3--18, 2006.

\bibitem{White2006}
F.~M. White and I.~Corfield.
\newblock {\em {Viscous fluid flow}}.
\newblock McGraw-Hill, New York, second edition, 2006.

\bibitem{Willis2006}
D.~J. Willis.
\newblock {\em {An Unsteady, Accelerated, High Order Panel Method with Vortex
  Particle Wakes}}.
\newblock Phd, Massachusetts Institute of Technology, 2006.

\bibitem{Willis2007}
D.~J. Willis, J.~Peraire, and J.~K. White.
\newblock {A combined pFFT-multipole tree code, unsteady panel method with
  vortex particle wakes}.
\newblock {\em International Journal for Numerical Methods in Fluids},
  53:1399--1422, 2007.

\bibitem{Zhu2007}
Q.~Zhu.
\newblock {Numerical Simulation of a Flapping Foil with Chordwise or Spanwise
  Flexibility}.
\newblock {\em AIAA Journal}, 45(10):2448--2457, 2007.

\bibitem{Zhu2008}
Q.~Zhu and K.~Shoele.
\newblock {Propulsion performance of a skeleton-strengthened fin.}
\newblock {\em Journal of Experimental Biology}, 211:2087--2100, 2008.

\bibitem{Zhu2002}
Q.~Zhu, M.~J. Wolfgang, D.~K.~P. Yue, and M.~S. Triantafyllou.
\newblock {Three-dimensional flow structures and vorticity control in fish-like
  swimming}.
\newblock {\em Journal of Fluid Mechanics}, 468:1--28, 2002.

\end{thebibliography}

\end{document}